\documentclass[12pt,a4paper]{article}
\usepackage[utf8]{inputenc}
\usepackage[dvipsnames]{xcolor}
\usepackage[normalem]{ulem}
\usepackage[bb=boondox]{mathalfa}
\usepackage[export]{adjustbox}
\usepackage{mathtools,resizegather,amsmath,amsfonts,amssymb,mathrsfs,amsthm,amsbsy,multirow,multicol,subcaption,dcolumn,makecell,scalerel,caption,jheppub,verbatim,braket,slashed, graphicx, physics,bbm,bm,algpseudocode, hyperref, algorithm, upgreek, xfrac}

\def\be{\begin{equation}}
\def\ee{\end{equation}}

\title{\Large Complexity of Quadratic Quantum Chaos}

\author[a]{Pallab Basu\,\href{https://orcid.org/0000-0003-0006-7240}
{\includegraphics[scale=0.05]{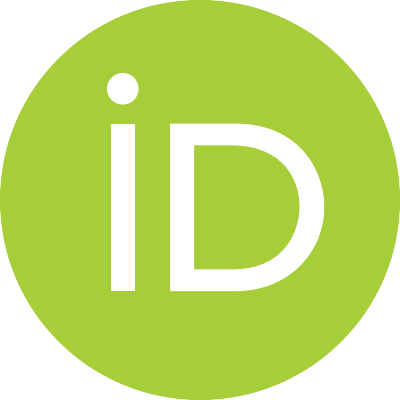}}\,,}
\author[a]{Suman Das\,\href{https://orcid.org/0000-0002-0053-3187}
{\includegraphics[scale=0.05]{orcidid.pdf}}\,,}
\author[b,c,d]{and Pratik Nandy\,\href{https://orcid.org/0000-0001-5383-2458}
{\includegraphics[scale=0.05]{orcidid.pdf}}\,}

\emailAdd{pallab.basu@wits.ac.za}
\emailAdd{suman.das@wits.ac.za}
\emailAdd{pratik.nandy@vub.be}

\affiliation[a]{Mandelstam Institute for Theoretical Physics, School of Physics,\\ University of the Witwatersrand, Johannesburg, WITS 2050, South Africa.}
\affiliation[b]{Theoretische Natuurkunde,\\ Vrije Universiteit Brussel (VUB) and International Solvay Institutes,\\ Pleinlaan 2, B-1050 Brussels, Belgium}
\affiliation[c]{Center for Gravitational Physics and Quantum Information,\\
Yukawa Institute for Theoretical Physics, Kyoto University,\\
Kitashirakawa Oiwakecho, Sakyo-ku, Kyoto 606-8502, Japan}
\affiliation[d]{Division of Fundamental Mathematical Science,\\
RIKEN Center for Interdisciplinary Theoretical and Mathematical Sciences (iTHEMS),\\
Wako, Saitama 351-0198, Japan}

\abstract{We investigate minimal two-body Hamiltonians with random interactions that generate spectra resembling those of Gaussian random matrices, a phenomenon we term \emph{quadratic quantum chaos}. Unlike integrable two-body fermionic systems, the corresponding hard-core boson models exhibit genuinely chaotic dynamics, closely paralleling the Sachdev–Ye–Kitaev (SYK) model in its spin representation. This chaotic behavior is diagnosed through spectral statistics and measures of operator growth, including Krylov complexity and the late-time decay of higher-order out-of-time-ordered correlators (OTOCs); the latter reveals the emergence of freeness in the sense of free probability. Moreover, the fractal dimension and Stabilizer R\'enyi entropy of a representative mid-spectrum eigenstate show finite-size deviations yet converge toward Haar-randomness as the system size increases. This convergence, constrained by local interactions, highlights the ``weakly chaotic’’ character of these eigenstates. Owing to their simplicity and bosonic nature, these minimal models may constitute promising and resource-efficient candidates for probing quantum chaos and information scrambling on near-term quantum devices.}

\begin{document}

\maketitle
\flushbottom

\section{Introduction}

Quantum chaos \cite{stockmann1999quantum} addresses the emergence of chaotic behavior in quantum systems, where the absence of classical trajectories demands novel diagnostics. Classical chaos arises in deterministic nonlinear dynamical systems whose evolution shows extreme sensitivity to initial conditions, often known as the butterfly effect. Nearby trajectories in phase space diverge exponentially at a rate set by the Lyapunov exponent \cite{strogatz2015nonlinear}, making long-term prediction effectively impossible despite deterministic temporal evolution. Chaotic motion is typically associated with mixing, ergodicity, and a complex structure of the underlying phase space, such as strange attractors \cite{strogatz2015nonlinear}. Well-known examples include the double pendulum, the planetary three-body problem, and turbulent flows. These features, however, rely on the notion of well-defined trajectories in phase space, a concept absent in quantum mechanics.

In quantum systems, chaos manifests both in universal features of the energy spectrum---well described by random matrix theory (RMT) \cite{Wigner_1951, Dyson1962a, mehta1991random, Cotler:2016fpe}---and in the scrambling of quantum information \cite{Hayden:2007cs}, which can be probed dynamically by out-of-time-order correlators (OTOCs) \cite{Larkin1969Quasiclassical, Maldacena:2015waa, Cotler:2017jue} and by a ``web of complexity measures'' such as operator entanglements \cite{Kudler-Flam:2019kxq} and Krylov space quantities \cite{Parker:2018yvk, Balasubramanian:2022tpr, Nandy:2024evd, Baiguera:2025dkc, Rabinovici:2025otw}. Unlike integrable systems, which exhibit Poissonian level statistics \cite{berry77}, the eigenspectra of chaotic systems show level repulsion and follow Wigner–Dyson distributions \cite{BGSconjecture}. The early-time exponential growth of OTOCs defines a quantum Lyapunov exponent $\lambda_L$, bounded by $\lambda_L \leq 2\pi k_B T/\hbar$ in thermal systems with a semiclassical gravity dual \cite{Maldacena:2015waa}.

The Sachdev--Ye--Kitaev (SYK) \cite{SachdevYe_1992, Kittu} model describes $N$ Majorana fermions with all-to-all random $q$-body interactions, described by the Hamiltonian:
\begin{align}
H = \sum_{1 \leq i_1 < i_2 < \cdots < i_q \leq N} J_{i_1 i_2 \cdots i_q}\, \chi_{i_1} \chi_{i_2} \cdots \chi_{i_q}\,, \label{sykham}
\end{align}
where the $J_{i_1 \ldots i_q}$ are all-to-all random couplings drawn from a Gaussian ensemble with zero mean and some specified variance. The SYK model provides a rare solvable setting that saturates the chaos bound \cite{Maldacena:2015waa}. In the large-$N$ limit, the SYK model reproduces RMT spectral statistics, exhibits maximal scrambling, and realizes features of near-AdS$_2$ holography \cite{Maldacena:2016hyu}. Its low-energy dynamics are governed by an emergent reparametrization symmetry weakly broken to $\mathrm{SL}(2,\mathbb{R})$, leading to a Schwarzian effective action that captures the leading chaotic behavior. These properties make SYK and its generalizations paradigmatic models for studying strongly correlated quantum matter and quantum gravity in low dimensions \cite{Chowdhury:2021qpy}. Moreover, recent studies have taken promising steps toward simulating the SYK model and its real-time dynamics on various quantum platforms, including trapped ions, superconducting circuits, and nuclear magnetic resonance quantum simulators \cite{Garcia-Alvarez:2016wem, Luo:2017bno}.

However, simulating the SYK model poses significant challenges for two main reasons. First, the Hamiltonian is highly dense: in the case of $q$-body interactions, the number of terms grows as $O(N^q)$ where $N$ denotes the total number of Majorana fermions in the system. This rapid scaling makes the numerical simulations increasingly difficult as the system size increases. Second, the inherently fermionic nature of the model introduces additional complexity. The Majorana fermions are mapped to spin degrees of freedom via the Jordan–Wigner (JW) transformation:
\begin{align}
    \chi_{2a-1} = (\sigma^z)^{\otimes (a-1)} \otimes \sigma^x_a \otimes (\sigma^z)^{\otimes (N_{\mathrm{Spin}}-a)}\,,~~\chi_{2a} = (\sigma^z)^{\otimes (a-1)} \otimes \sigma^y_a \otimes (\sigma_z)^{\otimes (N_{\mathrm{Spin}}-a)}\,, \label{JW1}
\end{align}
with $a = 1,2,\cdots, N_{\mathrm{Spin}}$, and $N_{\mathrm{Spin}} = N/2$. Here $\sigma^{x,y,z}_a$ are the two-dimensional Pauli matrices acting at site $a$. This leads to severe non-locality, hindering large-scale numerical simulations. This naturally motivates a fundamental question: can one construct a simplified version or a fermion-free surrogate that preserves the essential chaotic features of the SYK model, including RMT statistics, maximal scrambling \cite{Maldacena:2016hyu}, and holographic emergence, while remaining amenable to efficient classical or hybrid quantum–classical simulation \cite{Araz:2024xkw}?

To address this question, novel simplification strategies have been proposed. One approach is to sparsify the model, giving rise to the so-called \emph{sparse} SYK model \cite{Xu:2020shn}. In this variant, a fraction of the interaction terms is randomly deleted from the Hamiltonian, with the sparsity parameter $p$ controlling the deletion probability. As a result, for $q$-body interactions, the number of effective terms in the Hamiltonian scales as $O(p N^q)$, thereby substantially reducing the computational cost. The fully dense SYK model \eqref{sykham} corresponds to $p = 1$. Remarkably, this parameter admits a critical value $p_c$, which scales inversely with system size (with appropriate exponents) \cite{Garcia-Garcia:2020cdo, Orman:2024mpw, Nandy:2024wwv} such that, for $p > p_c$, the model continues to exhibit chaotic features, whereas for  $p < p_c$, it exhibits signatures of integrability due to emergent symmetries. In addition, the Gaussian-distributed random couplings of the original SYK Hamiltonian \eqref{sykham} can be replaced by binary variables \cite{Tezuka:2022mrr}, offering a further simplification without destroying the chaotic characteristics of the model.

To address the second challenge---namely, the non-locality of fermionic operators in the spin representation---recent works have proposed studying bosonic versions constructed directly from local spin operators \cite{Rosa:2019jin, Hanada:2023rkf, Swingle:2023nvv}. In this approach, the Jordan–Wigner strings are bypassed by defining a local operator basis built from Pauli matrices. Concretely, one introduces
\begin{align}
O_{2a-1} = \mathbb{I}^{\otimes (a-1)} \otimes \sigma^x_a \otimes \mathbb{I}^{\otimes (N_{\mathrm{Spin}}-a)}\,,~~~~ O_{2a} = \mathbb{I}^{\otimes (a-1)} \otimes \sigma^y_a \otimes \mathbb{I}^{\otimes (N_{\mathrm{Spin}}-a)}\,, \label{opp}
\end{align}
with $a = 1, 2, \cdots, N_{\mathrm{Spin}}$. The Pauli matrices $\sigma^{x}$ and $\sigma^{y}$ on the $a$-th lattice site, while acting as the identity on all other sites. This resembles the Jordan–Wigner transformation, except that the string of $\sigma^z$ matrices is replaced by identities. Consequently, the Clifford algebra structure is explicitly broken, and the fundamental constituents behave as hard-core bosons or spins. Using these operators, one can define a spin analogue of the SYK model---referred to as the Spin-SYK Hamiltonian (see Eq.\,\eqref{SpinSYKq})---which retains an SYK-like structure. Remarkably, even the Spin-SYK model \cite{Hanada:2023rkf}, which utilizes only two Pauli matrices, exhibits chaotic behavior for $q=4$ and is expected to flow to the same SYK-like fixed point for sufficiently large values of $q$ \cite{Swingle:2023nvv, Watanabe:2023vzo}.

Although large-$q$ models are often more naturally suited to holographic analyses, here we focus on two-body, or ``quadratic'', Spin-SYK models. Our aim is to investigate one of the simplest realizations of large-$N$ chaos. It is noteworthy that the quadratic structure of a model does not by itself ensure integrability or rule out chaotic dynamics; these features are dictated by the underlying operator algebra. Indeed, even nearest-neighbor quantum spin chains built from Pauli matrices are known to exhibit chaotic behavior \cite{Gubin_2012}.

While $q=2$ models are not anticipated to possess fixed points exhibiting emergent gravitational behavior \cite{Swingle:2023nvv}, they nevertheless present compelling opportunities for investigating diverse complexity measures through spectral statistics and operator growth. Such investigations provide valuable comparative insights when contrasted with the $q\geq 4$ regime.

Our investigation reveals several noteworthy findings:
\begin{itemize}
    \item \textbf{Chaotic behavior in quadratic interactions:} Contrary to naive expectations, even the $q=2$ Spin-SYK model, constructed using only two Pauli matrices per site, exhibits chaotic dynamics.\footnote{In contrast with the integrable fermionic SYK$_2$ model.} We referred to this as \emph{quadratic quantum chaos}. The short- and long-range spectral statistics show behavior consistent with the RMT, indicating the chaotic nature of the Hamiltonian. Further analysis addressed even simplified variants of the quartic model, which were also found to exhibit chaos (see Appendix \ref{app:simpham}).

 \item \textbf{Krylov space approach:} For an initial local operator, the quadratic Hamiltonian exhibits linear growth of Lanczos coefficients---features typically associated with chaotic Hamiltonians \cite{Parker:2018yvk}. Because our Hamiltonian has a spin reversal invariance, the Krylov space decomposes into even and odd sectors. As a result, the effective Krylov space dimensions are reduced to half of the theoretical expectation. Nevertheless, we often find that operators exhibit Krylov dimensions slightly smaller than this reduced value, indicating additional constraints on the accessible Hilbert space that merit further investigation.
    
\item \textbf{Freeness at late times:} In our quadratic model, we observe the emergence of freeness \cite{mingo} at late times, even at finite size: as local operators evolve under the Hamiltonian, they become statistically independent of their initial configurations \cite{Chen:2024zfj, Camargo:2025zxr, Jahnke:2025exd}. Within the realm of free probability theory, this late-time freeness provides a natural language for describing quantum ergodicity, offering a compact way to characterize the loss of initial memory and operator delocalization in chaotic many-body systems.

\item \textbf{Fractal dimension and the non-stabilizerness:} The fractal dimension \cite{IvanMultifractalRMT} and the Stabilizer R\'enyi entropy (SRE) \cite{Leone:2021rzd} of mid-spectrum eigenstates exhibit systematic deviations from Haar-random predictions. These deviations reveal that even mid-spectrum states remain short of full ergodicity, characterizing the mid-spectrum eigenstates of Spin SYK models as only ``weakly ergodic''---approaching Haar-randomness only in the infinite-size limit. Importantly, this behavior differs qualitatively from that of integrable systems, where the SRE remains well below the Haar-random benchmark, underscoring that the observed dynamics are closer to chaotic rather than integrable systems.

\end{itemize}

The paper is organized as follows. In Section \ref{sec:model}, we introduce the Spin-SYK Hamiltonians, with particular emphasis on the quadratic case ($q=2$). Section \ref{sec:eigen} presents an analysis of the eigenspectra of Spin-SYK models, benchmarked against RMT. As diagnostic probes, we employ level correlations, the spectral form factor (SFF), and Krylov spread complexity. Section \ref{sec:ogr} explores operator growth, including operator Krylov complexity, higher-order OTOCs, and the emergence of freeness. Section \ref{sec:eigprop} examines the eigenstate properties, namely the fractal dimensions and the SRE of a typical mid-spectrum eigenstate, and compares the results with Haar-random predictions. Section \ref{sec:con} concludes with a summary and possible future directions. Additional results for the $q \geq 2$ and some related simplified models are presented in the Appendices.

\section{The Quadratic Spin SYK Models}\label{sec:model}

The Hamiltonian of a spin SYK model is defined by random couplings among quantum $SU(2)$ spin operators on $N_{\mathrm{Spin}}$ sites. Hence, for $N$ Majorana fermions, we have $N_{\mathrm{Spin}} = N/2$. To this end, we rewrite the constituent spin operators from \eqref{opp}:
\begin{align}
O_{2a-1} = \mathbb{I}^{\otimes (a-1)} \otimes \sigma^x_a \otimes \mathbb{I}^{\otimes (N_{\mathrm{Spin}}-a)}\,,~~~~ O_{2a} = \mathbb{I}^{\otimes (a-1)} \otimes \sigma^y_a \otimes \mathbb{I}^{\otimes (N_{\mathrm{Spin}}-a)}\,, 
\end{align}
with $a = 1, 2, \cdots, N_{\mathrm{Spin}}$. Note that the operators with odd indices contain the $\sigma^x$, while even indices contain $\sigma^y$.

Using these operators, the Spin-SYK$_q$ Hamiltonian \cite{Hanada:2023rkf,Swingle:2023nvv} can be expressed in the SYK-like form:
\begin{align}
H_{\mathrm{Spin\,SYK}_q} = \sqrt{\frac{(q-1)!}{(2N_{\mathrm{Spin}})^{q-1}}} \sum_{1 \leq i_1 < i_2 < \ldots < i_q \leq 2N_{\mathrm{Spin}}} i^{\eta_{i_1 i_2 \ldots i_q}}\,  J_{i_1 i_2 \ldots i_q} \,  O_{i_1} O_{i_2} \ldots O_{i_q}\,, \label{SpinSYKq}
\end{align}
where $q$ denotes the number of spin operators in each interaction term. Here $\eta_{i_1, i_2,\ldots,i_q}$ is a function of the index configuration and takes values in $\in \{0, 1\}$, ensuring the Hermiticity of the Hamiltonian. As an illustrative example, consider the case $q=2$, the product $O_1 O_2$ produces a single $z$-string: $O_1 O_2 = i \sigma^z \otimes \ldots \otimes \mathbb{I}$, which is non-Hermitian. To render the Hamiltonian Hermitian, we instead include the term $i O_1 O_2$. In this case, we assign $\eta_{12} = 1$.

In the following, we will primarily focus on the $q=2$ Spin-SYK$_2$ case. The Hamiltonian of the model takes the form  
\begin{align}
H_{\mathrm{Spin\,SYK}_2} = \sqrt{\frac{1}{2N_{\mathrm{Spin}}}}  \sum_{1 \leq i_1 < i_2 \leq 2 N_{\mathrm{Spin}}} i^{\eta_{i_1 i_2}}\, J_{i_1 i_2} \,  O_{i_1} O_{i_2}\,. \label{SpinSYKq2}
\end{align}
However, when $a$ is odd, the combinations $i O_a O_{a+1}$ introduce a single $z$-string, which effectively acts as a self-site interaction. Consequently, this term does not represent a \emph{genuine} two-body interaction. By contrast, the remaining terms in the Hamiltonian are true two-body interactions. To isolate \emph{genuine} two-body interactions—that is, those involving operators acting on two distinct sites—it is often desirable to remove such self-site contributions. In practice, this amounts to deleting operator combinations where multiple components act on the same site.\footnote{This procedure can be naturally generalized to $q$-body interactions.} The resulting version of the model, in which all self-site interactions are excluded \cite{supp}, contains only \emph{genuine} two-body couplings; we refer to it as the ``genuine'' or gSpin-SYK model. For convenience, we use the notation (g)Spin-SYK to collectively denote both variants. The couplings $J_{i_1 i_2}$ or generally $J_{i_1 \ldots i_q}$ for general $q$-body interactions are sampled from a Gaussian distribution with zero mean and unit variance: $J_{i_1 \ldots i_q} \sim \mathcal{N}(0,1)$.

Despite its two-body structure, this model already exhibits hallmarks of quantum chaos, as revealed by its spectral statistics, operator growth, and eigenstate properties. In Appendix \ref{app:simpham}, we discuss an even simpler version of the quadratic model, which exhibits chaotic characteristics, such as eigenvalue repulsion.

We define the family of global operators
\begin{align}
\label{eq:concharge}
\Gamma^3 = \bigotimes_{i=1}^{N_{\mathrm{Spin}}} \sigma_i^{z} \, ,
\end{align}
This anticommutes with any of the constituent operators in Eq.\,\eqref{opp}: $\{\Gamma^3, O_k\} = 0$. Consequently, the Hamiltonian $H_{\mathrm{Spin\,SYK}_2}$ admits $\Gamma^3$ as a conserved $\mathbb{Z}_2$ charge, \emph{i.e.},
\begin{align}\label{eq:conQ}
[H_{\mathrm{Spin\,SYK}_2}, \Gamma^3] = 0 \, .
\end{align}
The operator $\Gamma^3$ has eigenvalues $\pm 1$, which label the two parity sectors with respect to the $\sigma^3$ basis. Consequently, the Hilbert space decomposes into even and odd subsectors, and the spectrum of $H_{\mathrm{Spin\,SYK}_2}$ splits accordingly.

\section{Eigenspectrum analysis of the Spin SYK models} \label{sec:eigen}

\subsection{Short-range spectral statistics: level spacing ratios}

\begin{figure}
    \hspace*{-0.3 cm}    \includegraphics[width=1\linewidth]{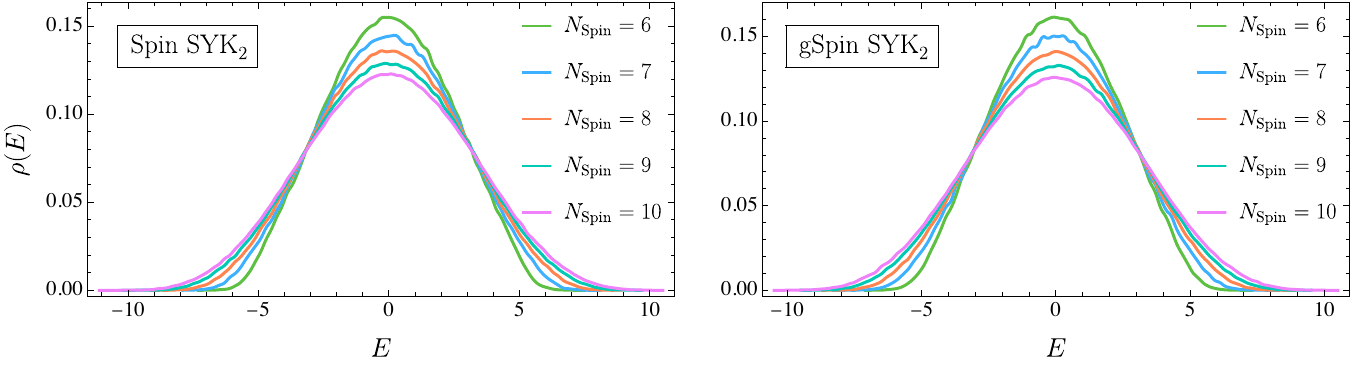}
    \caption{The Density of States (DOS) of the (g)Spin SYK models for various $N_{\mathrm{Spin}}$. We choose one parity sector of the Hamiltonian of dimension $d = 2^{N_{\mathrm{Spin}}-1}$. For each $N_{\mathrm{Spin}}$, total $2^{20-N_{\mathrm{Spin}}}$ ensemble of Hamiltonians are taken.}
    \label{fig:SYK2DOS}
\end{figure}

In this section, we examine the eigenspectrum and spectral diagnostics of (g)Spin SYK$_2$ models. Figure \ref{fig:SYK2DOS} shows the density of states (DOS) for the Spin SYK$_2$ model (left) and the gSpin SYK$_2$ model (right) for various $N_{\mathrm{Spin}}$. The Gaussian-like behavior of DOS is also common in the Majorana version \cite{Garcia-Garcia:2016mno}. The DOS for the Spin SYK$_4$ model is examined in Ref.\,\cite{Hanada:2023rkf}.

In quantum many-body systems with finite Hilbert spaces, spectral statistics provide a natural probe of quantum chaos. In chaotic systems, eigenvalues exhibit level repulsion, whereas in integrable systems, levels tend to cluster more randomly. These correlations manifest over short, medium, or long ranges, reflecting interactions among eigenvalues across the spectrum. One of the most widely used indicators of spectral correlations is the distribution of nearest-neighbor level spacings and their ratios, constructed from the ordered eigenvalues $\lambda_i$ of the Hamiltonian. These quantities capture the local correlations in the spectrum. However, in chaotic systems, spectral correlations are known to persist over long ranges \cite{Cotler:2016fpe}. To probe these long-range structures, one can examine higher-order spacing ratios, which quantify correlations between non-adjacent levels and thus offer a more refined diagnostic of spectral rigidity. The non-overlapping $k$-th order spacing ratio is defined as \cite{Oganesyan:2007wpd, Atas2013distribution, Tekur:2018nme}:
\begin{align}
     r_n^{(k)} = \min \left(\frac{s_{n+k}^{(k)}}{s_n^{(k)}}, \frac{s_n^{(k)}}{s_{n+k}^{(k)}}\right)\,, \quad \text{with} \quad s_n^{(k)} = \lambda_{n+k} - \lambda_n\,,
\end{align}
where $s_n^{(k)}$ denotes the energy difference between levels separated by $k$ positions. For $k=1$, this reduces to the well-known nearest-neighbor ratio \cite{Oganesyan:2007wpd, Atas2013distribution}. These spacing ratios possess desirable features: they are dimensionless, free from the need for spectral unfolding, and independent of the local density of states. As $k$ increases, the $r_n^{(k)}$ provide greater sensitivity to long-range spectral correlations, making them particularly valuable for detecting subtle spectral features and phase transitions.

In quantum chaotic systems described by RMT, the distributions of $r_n^{(k)}$ exhibit universal behavior, though the precise functional form and peak location vary with $k$. For Gaussian ensembles labeled by Dyson index $\beta = 1, 2, 4$ (corresponding to GOE, GUE, and GSE, respectively), the distribution of the nearest-neighbor ratio $r \equiv r_n^{(1)}$ takes the compact analytic form \cite{Atas2013distribution}:
\begin{align}
   P^{(1)}_{\beta}(r) = Z_{\beta} \, \frac{(r + r^2)^\beta}{(1 + r + r^2)^{1 + \frac{3}{2}\beta}}\,, \label{pbeta}
\end{align}
where $Z_\beta$ ensures normalization $\int_0^1 P_{\beta}(r) \, dr = 1$.

Remarkably, the higher-order spacing ratio distributions can also be related to the universal form \eqref{pbeta} through an effective Dyson index $\beta_{\mathrm{eff}}$ \cite{Tekur:2018nme}:
\begin{align}
    P^{(k)}_{\beta}(r) = P^{(1)}_{\beta_{\mathrm{eff}}(k)}(r)\,,~~~~ \beta_{\mathrm{eff}}(k) = \frac{k(k+1)}{2} \beta + (k-1)\,,~~ k \geq 1\,.
\end{align}
The higher-order ratios reflect the increasing rigidity of the spectrum as one probes longer-range correlations. For Gaussian ensembles, the resulting $P^{(k)}_{\beta}(r)$ becomes increasingly peaked with growing $k$, signifying enhanced correlations across energy levels. In contrast, integrable or many-body localized systems, which follow Poisson statistics, show $r$-distributions that remain nearly flat and $k$-independent, indicating the absence of such extended and long-range level repulsion.\footnote{For Poisson statistics, the probability distribution of the higher-order ratios is given by \cite{Tekur:2024kyt}
\begin{align}
    P^{(k)} (r) = \frac{(2k-1)!}{[(k-1)!]^2} \frac{2 r^{k-1}}{(1+r)^{2k}}\,.
\end{align}}

\begin{figure}
    \hspace*{-0.3 cm}
\includegraphics[width=1\linewidth]{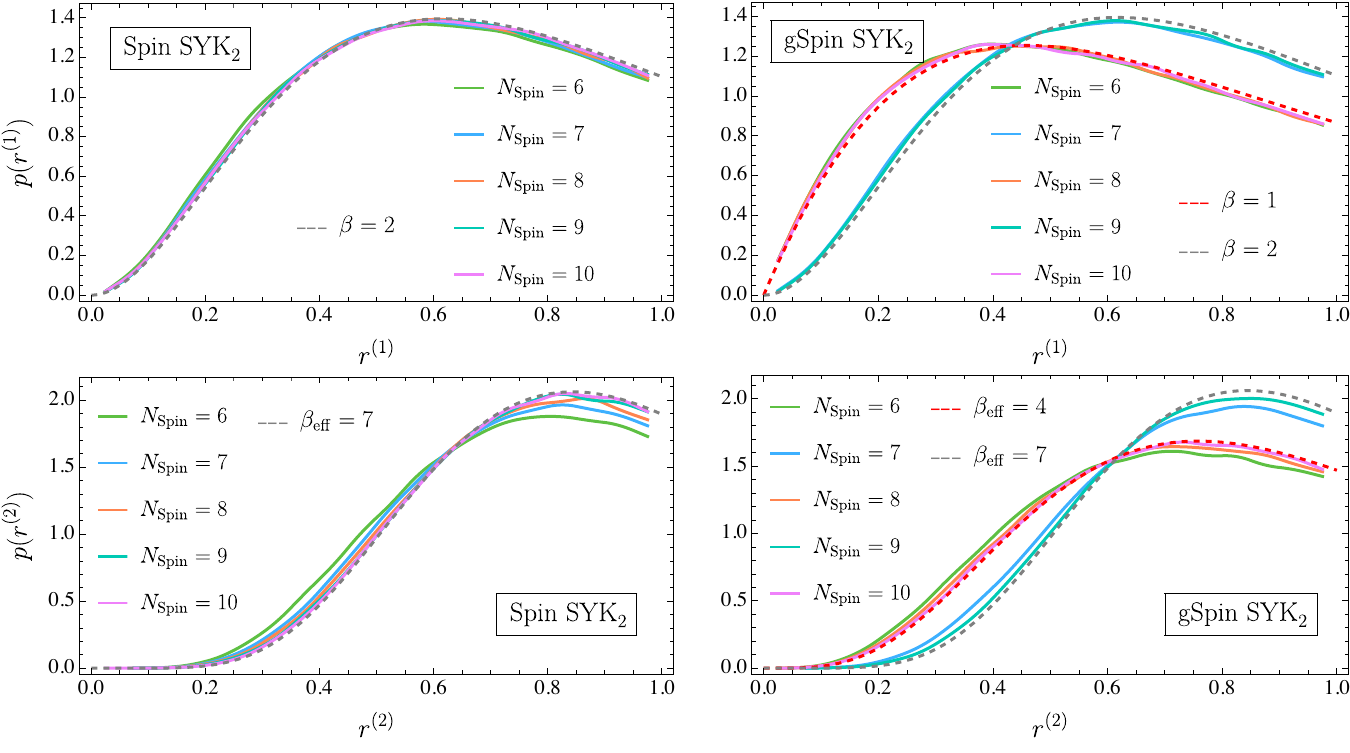}
    \caption{The nearest ($k = 1$, top panel) and next-to-nearest ($k = 2$, bottom panel) level spacing ratio distributions for the Spin SYK$_2$ (left) and (g)Spin SYK$_2$ (right) models for various $N$. Different $N$ correspond to the universal GUE class for SYK$_2$ model, while they oscillate between the GOE (even $N_{\mathrm{Spin}}$) and GUE (odd $N_{\mathrm{Spin}}$) class for the gSpin SYK$_2$ model. The system parameters are the same as Fig.\,\ref{fig:SYK2DOS}.}
    \label{fig:rratiok1k2SpinSYK}
\end{figure} 

Figure \ref{fig:rratiok1k2SpinSYK} illustrates the behavior of the nearest and next-to-nearest (left panel: Spin SYK$_2$ and right panel: gSpin SYK$_2$) level spacing ratio distributions. For $k = 1$, the spacing ratio distribution corresponds to the standard Gaussian RMT ensemble with Dyson index $\beta$, \emph{i.e.}, $\beta{\mathrm{eff}} = \beta$. For the Spin SYK$_2$ model, owing to the symmetry properties of the Hamiltonians, all values of $N_{\mathrm{Spin}}$ in these models belong to the universality class of the GUE \cite{Hanada:2023rkf}, corresponding to Dyson index $\beta = 2$. On the other hand, the spectra change between the GOE and GUE classes. For $k = 2$, the effective Dyson index becomes $\beta_{\mathrm{eff}} = 3\beta + 1$. For instance, in the case of the GOE with $\beta = 1$, this maps to an effective $\beta_{\mathrm{eff}} = 4$, which closely resembles the spectral statistics of the GSE. In Fig.\,\ref{fig:rratiok1k2SpinSYK}, the universality class of $\beta=1$ and $\beta=2$ is reflected in the $k=2$ statistics through the effective Dyson index $\beta{\mathrm{eff}} = 4$ and $\beta{\mathrm{eff}} = 7$, as shown in the bottom panels.

To quantify these distributions, it is often useful to consider the mean spacing ratio, defined as $\langle r^{(k)} \rangle = \int_0^1 r P^{(k)}_{\beta}(r) dr$. For the nearest-neighbor case ($k=1$), this yields the well-known values $\langle r^{(1)} \rangle \approx 0.53$, $0.60$, and $0.67$ for GOE, GUE, and GSE, respectively \cite{Atas2013distribution}. On the other hand, for $k = 2$ with $\beta_{\mathrm{eff}} = 4$ and $\beta_{\mathrm{eff}} = 7$, the bottom panels of Fig.\,\ref{fig:rratiok1k2SpinSYK} shows that the mean ratio increases to $\langle r^{(2)} \rangle \approx 0.65$ and $\langle r^{(2)} \rangle \approx 0.73$ respectively, corresponding to a narrower distribution with a more sharply defined peak, reflecting long-range spectral correlations. We note that similar chaotic features in the nearest-gap ratio were previously observed in a quadratic model in Ref.\,\cite{Rosa:2019jin} in the context of quantum batteries.

\subsection{Long-range spectral statistics: Spectral form factor and chaos} \label{sec:SFF}

Another widely used diagnostic of quantum chaos is the Spectral Form Factor (SFF), defined as the Fourier transform of the two-point correlation function of the energy spectrum \cite{Guhr:1997ve, Cotler:2016fpe}. The normalized SFF is defined as
\begin{align}
\mathrm{SFF}(t) = \bigg|\frac{Z(\beta_T + it)}{Z(\beta_T)}\bigg|^2 = \bigg|\frac{\mathrm{Tr}(e^{-(\beta_T + it)H})}{\mathrm{Tr}(e^{-\beta_T H})}\bigg|^2\,,
\end{align}
where $\beta_T = 1/T$ is the inverse temperature and $t$ denotes real time. Equivalently, the SFF can be interpreted as the survival (or return) probability of the coherent Gibbs state \cite{delCampo:2017bzr}. In practice, one typically computes the disorder-averaged SFF to extract universal spectral features.

The SFF is particularly useful for tracking the evolution of eigenvalue correlations across different time scales \cite{Brezin_Hikami_SFF, Guhr:1997ve, Cotler:2016fpe}. At early times $t \ll t_{\mathrm{Th}}$, the SFF exhibits a \textbf{dip} from its initial value $\mathrm{SFF}(0) = 1$, reflecting the onset of decoherence. In the intermediate regime $t_{\mathrm{Th}} \ll t \ll t_{\mathrm{H}}$, it displays a linear \textbf{ramp}: $\mathrm{SFF}(t) \approx t/t_{\mathrm{H}}$ \cite{Cotler:2016fpe}, which originates from universal long-range spectral correlations, a hallmark of quantum chaos. Here $t_{\mathrm{Th}}$ is the Thouless time \cite{Schiulaz:2018lwa}, associated with the Thouless energy \cite{THOULESS197493}: energy levels separated by less than this scale are typically correlated. The second scale is the Heisenberg time $t_{\mathrm{H}}$, set by the inverse mean level spacing. The slope of the ramp depends on the symmetry class (GOE, GUE, GSE) of the spectrum. Finally, at late times $t \gg t_{\mathrm{H}}$, the finite Hilbert space dimension forces the SFF to approach a \textbf{plateau} value of $Z(2\beta_T)/Z(\beta_T)^2$ \cite{Cotler:2016fpe}, \emph{i.e.}, the purity of the thermal density matrix.

\begin{figure}
    \hspace*{-0.3 cm}
\includegraphics[width=1\linewidth]{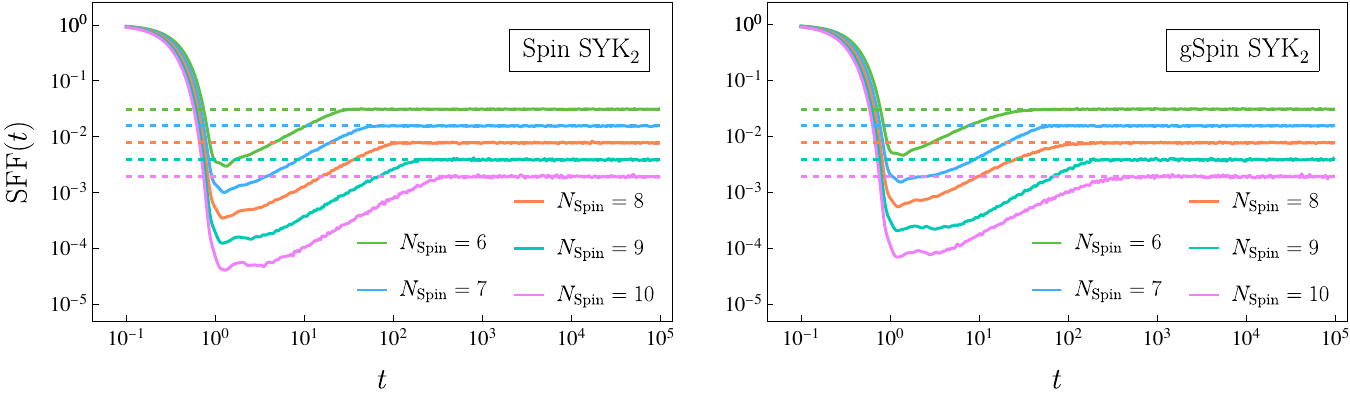}
    \caption{Time evolution of the Spectral Form Factor (SFF) for the (g)Spin SYK$_2$ models. Both models exhibit a clear ramp in the SFF, signaling chaotic behavior. The parameters are the same as in Fig.\,\ref{fig:SYK2DOS}.}
    \label{fig:SFFcompplotSpinSYK}
\end{figure}

\begin{figure}
    \hspace*{-0.3 cm}
    \includegraphics[width=1\linewidth]{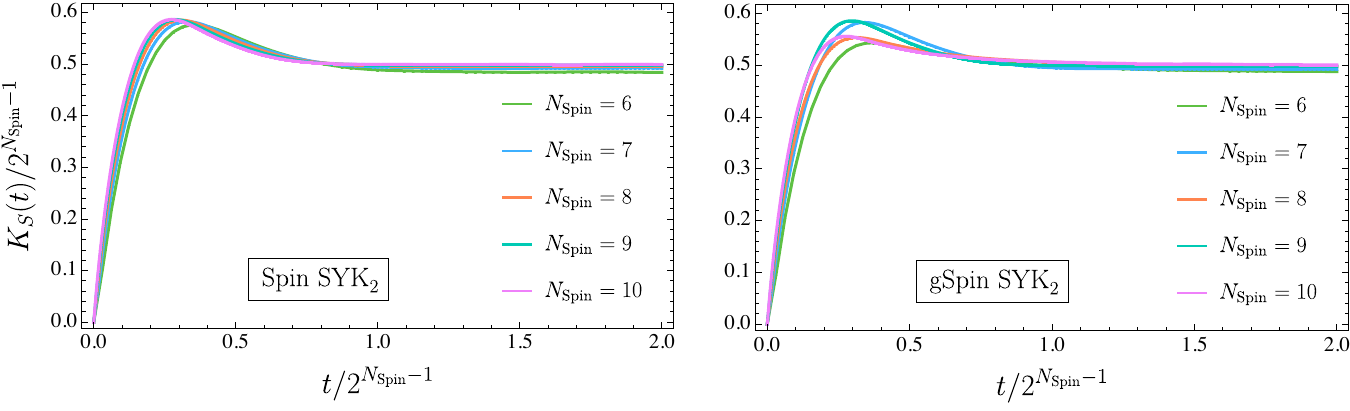}
    \caption{Time evolution of the Krylov spread complexity for the (g)Spin SYK$_2$ models. Both models exhibit a pronounced peak in complexity, signaling chaotic behavior. The height of the peak in the left panel is similar for all $N_{\mathrm{Spin}}$, while in the right panel, the height is different for different $N_{\mathrm{Spin}}$. The parameters are the same as in Fig.\,\ref{fig:SYK2DOS}.}
    \label{fig:kcompplotSpinSYK}
\end{figure}

Figure \ref{fig:SFFcompplotSpinSYK} shows the behavior of the SFF for the (g)Spin SYK$_2$ models for $\beta_T = 0$. Both models exhibit a clear ramp, signaling the presence of long-range spectral correlations typically associated with chaotic dynamics. In either case, the SFF saturates at $1/d = 2^{1-N_{\mathrm{Spin}}}$, \emph{i.e.}, the inverse dimension, where $d = 2^{N_{\mathbf{Spin}}-1}$ is the dimension of the one-parity sector of the Hamiltonian. The smoothness or sharp transition to the plateau is consistent with the observation in Ref.\,\cite{Cotler:2016fpe}.

The characteristic ramp in the spectral form factor (Fig.\,\ref{fig:SFFcompplotSpinSYK}) finds a parallel in the behavior of the Krylov (spread) complexity \cite{Balasubramanian:2022tpr}, normalized by the Hilbert space dimension, as shown in Fig.\,\ref{fig:kcompplotSpinSYK}. For the infinite-temperature Gibbs state taken as the initial state, the time evolution exhibits the standard \emph{growth–peak–slope–plateau} structure: an early-time linear growth, a pronounced intermediate peak, and a gradual approach to the late-time saturation value $(d-1)/(2d)$ \cite{Erdmenger:2023wjg}, irrespective of integrability (or lack thereof) of the system. In the left panel, the peak height is approximately independent of $N_{\mathrm{Spin}}$. However, for the \emph{genuine} two-body case, the underlying universality class, determined by $N_{\mathrm{Spin}}$, is either GOE or GUE, and manifests as distinct peak heights across different $N_{\mathrm{Spin}}$, as shown in the right panel. The presence of a peak is a strong indicator of correlations in the eigenspectra and is absent for integrable systems \cite{Balasubramanian:2022tpr}, which has motivated tracking it as a signature of chaotic–integrable transitions \cite{Camargo:2024deu, Bhattacharjee:2024yxj, Alishahiha:2024vbf} in RMT and many-body systems. The initial growth rate of the complexity is set by the characteristic energy scale of the Hamiltonian, which is determined by the variance of the couplings (or, equivalently, by an overall multiplicative normalization), in a manner similar to the disconnected part of the SFF \cite{Cotler:2016fpe}.

\section{Operator growth in the Spin SYK Models}\label{sec:ogr}
After analyzing the eigenspectrum, we turn to a more detailed investigation of the scrambling and chaotic properties of the model via operator growth. As expected, the growth behavior depends on the specific choice of probe operator. Nevertheless, for local operators that are not conserved quantities, the growth exhibits fairly universal features. In what follows, we focus primarily on the local operators:
\begin{align}
\begin{split}
        \mathsf{O}_1 &\equiv O_1 = \sigma_x \otimes \mathbb{I}  \otimes \mathbb{I} \otimes \cdots \otimes \mathbb{I} \,, \\
    \mathsf{O}_2 &\equiv O_1 O_4 = \sigma_x \otimes \sigma_y  \otimes \mathbb{I} \otimes \cdots \otimes \mathbb{I}\,,\\
     \mathsf{O}_3 &\equiv O_1 O_4 O_5 = \sigma_x \otimes \sigma_y \otimes \sigma_x \otimes \cdots \otimes \mathbb{I}\,.
    \label{inop1}
\end{split}
\end{align}
Note that for any even-spin Hamiltonian \eqref{SpinSYKq} possesses spin-reversal symmetry generated by the conserved charge $\Gamma^3$ (Eq.\,\eqref{eq:conQ}). Consequently, operators with an even number of spins do not mix with those containing an odd number under time evolution. The operators \eqref{inop1} provide examples of odd and even operators.

We examine the Lanczos spectrum \cite{viswanath1994recursion}, Krylov complexity, and cumulative higher-order OTOCs under the evolution of Spin-SYK models. In particular, the latter reveals the emergence of operator freeness at late times, a direct manifestation of the underlying chaotic dynamics. Our primary focus is on the growth of the $O_1$ operator, though the growth of other operators will also be highlighted.

\subsection{Lanczos spectra and Krylov complexity}

\begin{figure}
    \hspace*{-0.6 cm}
    \includegraphics[width=1.03\linewidth]{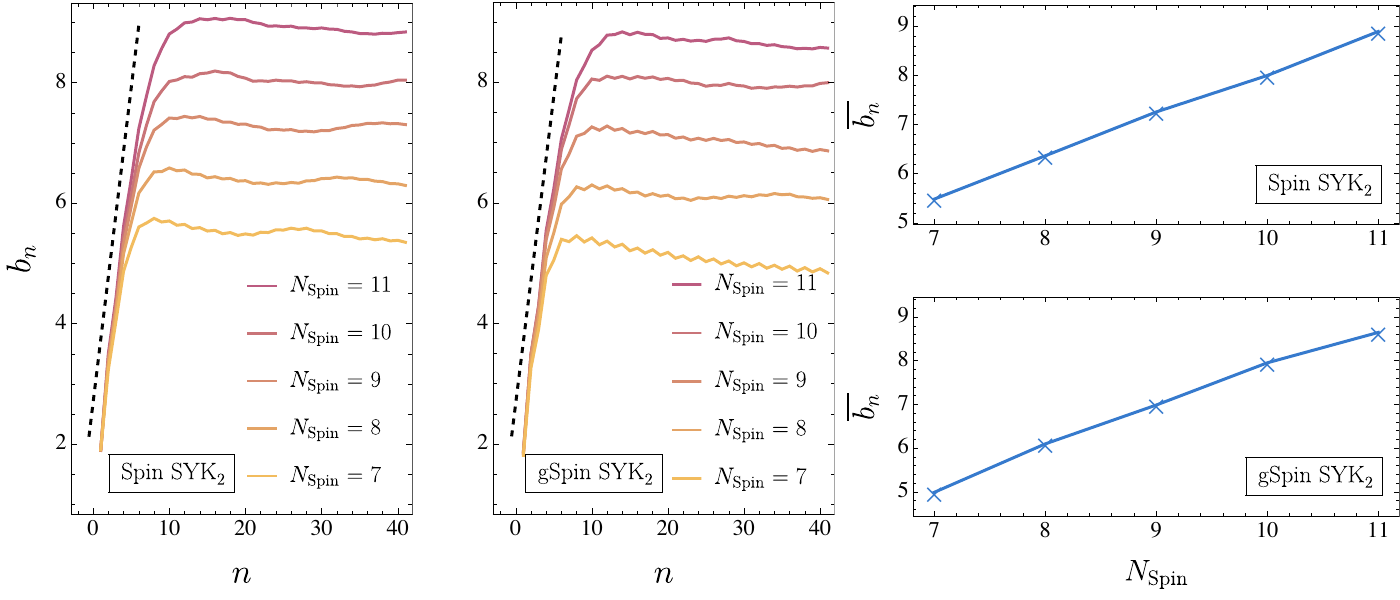}
    \caption{The behavior of Lanczos coefficients $b_n$ for Spin SYK$_2$ (left) and (g)Spin SYK$_2$ (middle) models, for five different system sizes. In both cases, the coefficients grow linearly before saturating to a plateau due to the finite system size. The black dashed lines are the fitted results with slope $m_{b_n} = 1.04$ (for Spin SYK$_2$) and $m_{b_n} = 0.98$ (for Spin SYK$_2$), respectively (slightly shifted for the visual interpretation), showing the linear growth of the Lanczos coefficients. The initial operator is $O_1$, and for each system size, averages were taken over $2^{16-N_{\mathrm{Spin}}}$ Hamiltonian realizations. Right: The finite-size effect of the plateau, which linearly increases with the system size $N_{\mathrm{Spin}}$.}
    \label{fig:plotbnggSpinSYK2}
\end{figure}

Following the method described in Appendix \ref{app:krycomplex}, we compute the Lanczos spectra for the (g)Spin SYK$_2$ models and evaluate the associated Krylov complexity. 

\begin{figure}
    \centering
\includegraphics[width=0.78\linewidth]{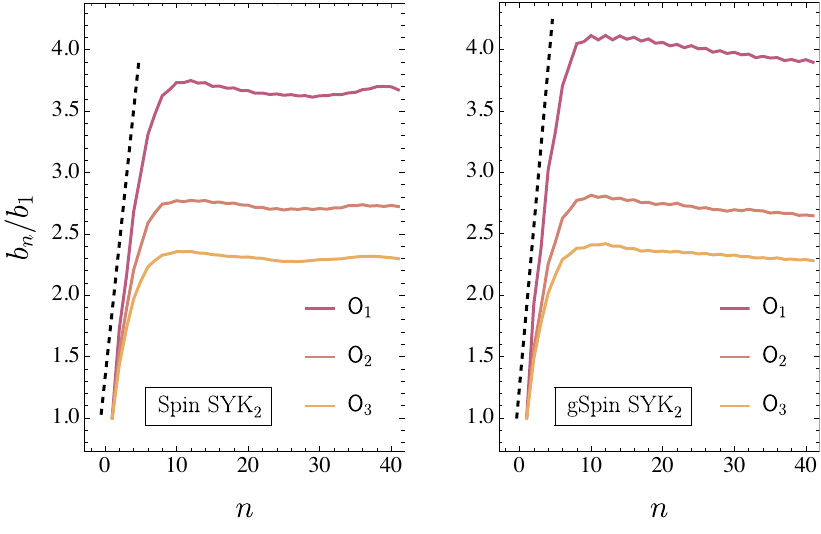}
    \caption{The Lanczos coefficients (scaled by the first coefficient) for different operator choices \eqref{inop1} for Spin SYK$_2$ (left) and gSpin SYK$_2$ model (right). The saturation for the gSpin SYK$_2$ model is higher because $b_1^{~\mathrm{Spin SYK_2}} \geq b_1^{~\mathrm{gSpin SYK_2}}$, as seen from Fig.\,\ref{fig:plotbnggSpinSYK2}. System size is $N_{\mathrm{Spin}} = 9$, and $100$ Hamiltonian realizations are taken.}
\label{fig:plotbnSpinSYK2diffO}
\end{figure}

Figure \ref{fig:plotbnggSpinSYK2} demonstrates the Lanczos coefficients $b_n$ for Spin SYK$_2$ (left panel) and gSpin SYK$_2$ (middle panel) with five different system sizes, $N_{\mathrm{Spin}} \in \{7, 8, 9, 10, 11\}$. For each system size, we average over $2^{16-N_{\mathrm{Spin}}}$ realizations of the Hamiltonian. The coefficients exhibit an initial linear growth regime (shown by the black dashed line), followed by saturation due to finite-size effects \cite{Jian:2020qpp, Nandy:2024evd}.\footnote{This saturation is only valid in the mid $n$ regime, since the coefficients ultimately vanish at the end of the Krylov space. Compare this result with the results from Majorana SYK$_4$ \cite{Jian:2020qpp, Nandy:2024evd}, or complex SYK$_4$ \cite{Rabinovici:2020ryf}.} The linear growth is a hallmark feature of the chaotic systems \cite{Parker:2018yvk}, except for some pathological examples \cite{Dymarsky:2021bjq, Bhattacharjee:2022vlt}. The saturation level is slightly lower for the gSpin SYK$_2$ models compared to the Spin SYK$_2$ case. The saturation value increases with system size as shown in the right panels, and in the thermodynamic limit, only the slope of $b_n$ becomes physically relevant. Since the dynamics are unitary, we find $a_n = 0$ throughout the entire Hamiltonian evolution. 
Analogously, the behavior of $b_n/b_1$ for three different choices of operator \eqref{inop1} in the Spin SYK$_2$ model is shown in Fig.\,\ref{fig:plotbnSpinSYK2diffO}.

\begin{figure}
    \hspace*{-0.3 cm}
\includegraphics[width=1\linewidth]{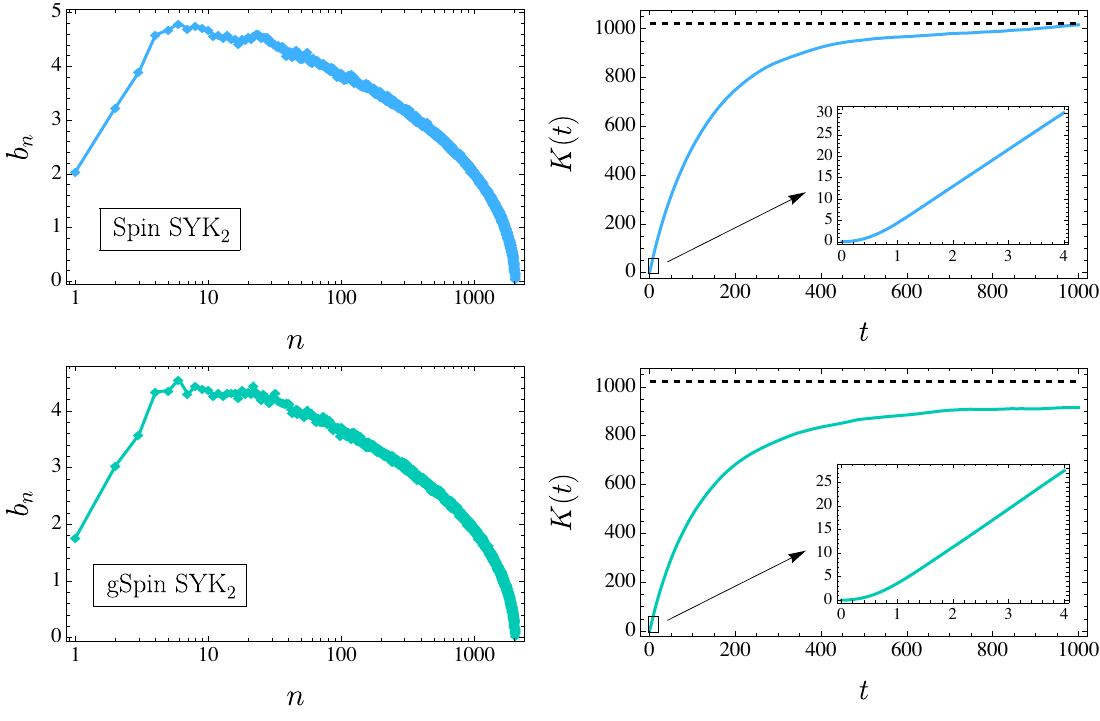}
\caption{Growth of Lanczos coefficients $b_n$ and Krylov complexity for Spin SYK$_2$ (upper panel) and gSpin SYK$_2$ (lower panel) for an odd operator $O_1$. The Krylov space dimension is $K_{\mathrm{dim}} = 2046 + o(1)$ for either case, where the $o(1)$ fluctuations may arise due to the randomness of the model. Correspondingly, the maximal saturation value of complexity is $K_{\mathrm{dim}}/2$, shown by the black dashed line in the right figures. The inset shows the early-time behavior. We take $N = 6$ with $20$ realizations of Hamiltonians.}
    \label{fig:bnKryplotggSpinSYK2N6}
\end{figure}

However, as seen in Fig.\,\ref{fig:plotbnggSpinSYK2}, the saturation does not persist indefinitely---the Lanczos coefficients eventually decrease and vanish at the end of the Krylov chain, which defines the dimension of the operator Krylov space, $K_{\mathrm{dim}}$. Figure \ref{fig:bnKryplotggSpinSYK2N6} shows the complete Lanczos spectrum for $N_{\mathrm{Spin}} = 6$ in both Spin SYK$_2$ (top left) and gSpin SYK$_2$ (bottom left) models, averaged over $20$ Hamiltonian realizations. 

For the operator $O_1$, the Krylov space dimension is $K_{\mathrm{dim}} \simeq 2046 + o(1)$. As expected, this is roughly half of the maximum possible dimension, $K_{\mathrm{dim}}^{\mathrm{max}} = d^2 - d + 1 = 4033$ (where $d = 2^6$) \cite{Rabinovici:2020ryf}. 
This indicates that the operator does not explore the maximal Krylov space, but is instead restricted to about half of it. The $o(1)$ fluctuations can arise from the randomness of the Hamiltonian, but vanish in the limit of large ensemble averages.

Correspondingly, the (operator) Krylov complexity exhibits an initial growth, followed by a linear regime and eventual saturation---a behavior typical of chaotic systems \cite{Parker:2018yvk, Barbon:2019wsy}. The inset displays a magnified view of the early-time behavior. The saturation value of the Krylov complexity is $K_{\mathrm{dim}}/2$ for the Spin SYK$_2$ model and slightly lower for the (g)Spin SYK$_2$ model. In both cases, the values are significantly below the maximal possible $K_{\mathrm{dim}}^{\mathrm{max}}/2$, again indicating that the operator in Eq.\,\eqref{inop1} does not explore the full Krylov space, consistent with the observation from Lanczos spectra. This behavior is reminiscent of spin chains with explicit integrability breaking \cite{Rabinovici:2022beu}. A similar order of magnitude of the Krylov dimension is also observed in the Spin-SYK$_4$ model (see Appendix \ref{AppB}). In all our calculations, Krylov complexity is computed separately for each random realization (\emph{i.e.}, for each Lanczos sequence obtained from a given Hamiltonian and operator), and the averaging is performed only at the final stage, namely, on the Krylov complexity $K(t)$ itself. Therefore, the Lanczos coefficients in Fig.\,\ref{fig:bnKryplotggSpinSYK2N6} (also Fig.\,\ref{fig:bnKryplotggSpinSYK4L6} in the Appendix) are not taken as input for the computation of Krylov complexity; they are separately extracted from the entire set of Lanczos coefficients to demonstrate the behavior of their overall growth. This ensures that all localization effects in Krylov space are correctly captured and that our results can be reliably interpreted.

\subsection{Cumulative OTOC and free probability}

In the context of quantum chaos, free probability theory \cite{voiculescu1986addition, voiculescu1987multiplication} furnishes a powerful analytic framework for characterizing the emergence of universal behavior in operator dynamics. Central to this framework is the concept of \emph{freeness} (or free independence), a non-commutative generalization of statistical independence \cite{mingo}. Two operators are defined to be \emph{free} if all their connected mixed cumulants vanish identically \cite{Pappalardi:2023nsj, Fritzsch:2024qjn}.  This notion of freeness formalizes the intuition \cite{Chen:2024zfj, Camargo:2025zxr, Jahnke:2025exd} that, at late times, a local operator evolving under a chaotic Hamiltonian becomes statistically independent or free (in the free probability sense) from its initial state.

Ideally, this free independence or freeness emerges rigorously in the limit of a large Hilbert space dimension $d \rightarrow \infty$, a property known as the \emph{asymptotic free independence} or simply \emph{asymptotic freeness}\footnote{This notion of asymptotic freeness should not be confused with the asymptotic freedom in gauge theories, most notably in quantum chromodynamics.} \cite{nica2006lectures}. Strikingly, recent numerical investigations \cite{Chen:2024zfj, Camargo:2025zxr, Jahnke:2025exd} demonstrate that signatures of freeness already manifest at finite, moderately large $d$. This observation suggests that chaotic dynamics efficiently drive local operator ensembles toward their universal free-probabilistic regime, well before the strict asymptotic limit is reached.

A systematic probe of emergent freeness is provided by the composite operator $X(t)=\mathcal{O}(0)+\mathcal{O}(t)$, where $\mathcal{O}(0)$ is a local operator at the initial time and $\mathcal{O}(t)$ its time-evolved image under chaotic dynamics. At early times, the spectral statistics of $X(t)$ follow the \emph{classical convolution} of the eigenvalue measures of $\mathcal{O}(0)$ and $\mathcal{O}(t)$, while at late times chaotic evolution drives them toward the \emph{free convolution} predicted by free probability theory. For instance, if $\mathcal{O}(0)$ is a spin-$1/2$ Pauli operator with eigenvalues ${\pm 1}$, the classical convolution yields a three-point distribution supported on $\{-2,0,+2\}$, whereas the onset of freeness implies that the spectral density of $X(t)$ converges to the universal arcsine law: $\rho = \frac{1}{\pi\sqrt{4-x^2}}$ with $x \in [-2,2]$, which is the canonical free-convolution outcome of two Bernoulli spectra \cite{Chen:2024zfj, Camargo:2025zxr, Jahnke:2025exd}.

\begin{figure}
    \hspace*{-0.3 cm}
\includegraphics[width=1\linewidth]{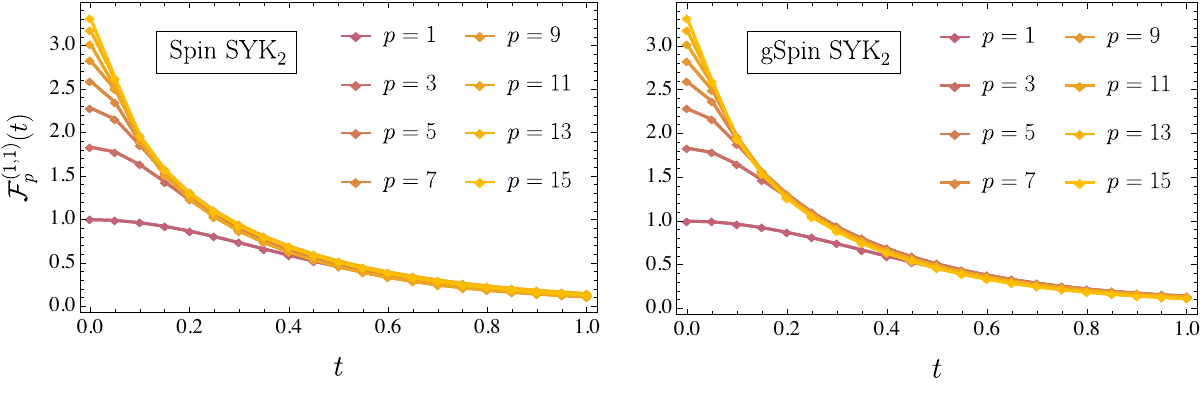}
    \caption{Cumulative OTOC for Spin SYK$_2$ (left) and gSpin SYK$_2$ (right) with the operator choice $O_\mathbb{I} = \sigma_x \otimes \mathbb{I} \otimes \cdots \otimes \mathbb{I}$ decays to zero at late times. Here $p$ is the number of terms considered in the sum \eqref{cumOTOC}. We take $N_{\mathrm{Spin}} = 9$ with $50$ realizations of Hamiltonians.}
    \label{fig:cummOTOCggSpinSYK}
\end{figure}

Although the full spectrum encodes complete information about the emergence of free independence at a given time $t$ and is directly linked to higher-order out-of-time-ordered correlators (OTOCs) \cite{Cotler:2017jue}, computing it at multiple times is often computationally intensive. In the context of freeness, higher-order OTOCs have been shown to play a central role in setting distinct dynamical timescales in quantum circuit models and ensembles of random unitaries \cite{Vallini:2024xab, Fritzsch:2025arx, Dowling:2025cxr}. Remarkably, much of this information---particularly the aspect relevant for diagnosing freeness---can be efficiently captured by the \emph{cumulative} OTOC \cite{Jahnke:2025exd}, defined as the partial sum of higher-order contributions:
\begin{align}
    \mathcal{F}_{p}^{(i,j)}(t) :=\sum_{n = 1}^{p} \frac{1}{n} \,\langle (\mathcal{O}_i(0) \mathcal{O}_i(t))^n \rangle^2 \,. \label{cumOTOC}
\end{align}
Here, $p$ denotes the number of terms retained in the sum, \emph{i.e.}, up to the $p$-th order OTOC.\footnote{The cumulative OTOC \eqref{cumOTOC} can be defined more generally with different choices of weight, without restricting to the harmonic weight. It was first introduced in Ref.\,\cite{Jahnke:2025exd}, where several weight factors—including the harmonic one—were considered. The harmonic weight, however, admits a natural geometric interpretation, as discussed in \cite{Shreyatalk}.} The operator $\mathcal{O}_i(0)$ acts locally on the $i$-th site (qubit) Hilbert space, tensor-producted with the identity on the remainder of the system. The factor $1/n$ serves as a weight to suppress contributions from higher-order OTOCs at each order $n$ \cite{Vallini:2024xab, Jahnke:2025exd}. As shown in Ref.\,\cite{Jahnke:2025exd}, chaotic dynamics drive the cumulative sum \eqref{cumOTOC} to decay at late times, vanishing exponentially with a rate determined by the two-point function.

\begin{figure}
    \hspace*{-0.3 cm}
\includegraphics[width=1\linewidth]{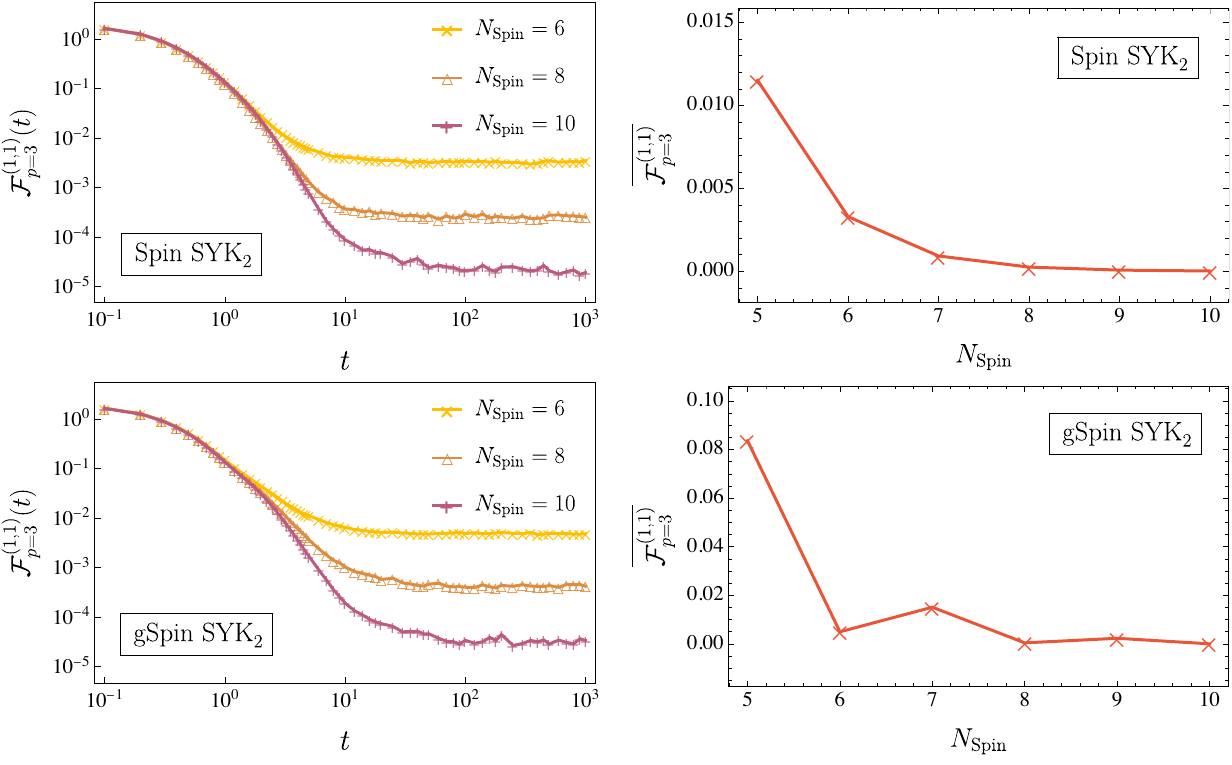}
    \caption{Finite-size analysis of cumulative OTOC for Spin SYK$_2$ (top panel) and gSpin SYK$_2$ (bottom panel) with the operator choice $O_\mathbb{I} = \sigma_x \otimes \mathbb{I} \otimes \cdots \otimes \mathbb{I}$ decays to zero at late times. Here $p = 3$ is the number of terms considered in the sum \eqref{cumOTOC}. The $t=0$ value corresponds to the Harmonic number $H_3 = \sum_{n=1}^3 (1/n) = 1.833$. The left panel shows the full profile of cumulative OTOC with three different choices of $N_{\mathrm{Spin}}$. The right panel shows that the saturation value decreases with $N_{\mathrm{Spin}}$, signaling the asymptotic freeness of the operator. Here, we take $2^{16-N_{\mathrm{Spin}}}$ realizations of Hamiltonians.}
\label{fig:finesizecummOTOCggSpinSYK}
\end{figure}

Figure \ref{fig:cummOTOCggSpinSYK} depicts the cumulative OTOC for the operator $O_1 = \sigma_x \otimes \mathbb{I} \otimes \cdots \otimes \mathbb{I}$ in the Spin SYK$_2$ (left panel) and gSpin SYK$_2$ (right panel) models, for a system of size $N_{\mathrm{Spin}}=9$ averaged over $50$ Hamiltonian realizations. The sum in \eqref{cumOTOC} is truncated to $p=15$ terms. In both cases, the cumulative OTOC exhibits a similar decay profile, vanishing at late times and closely mirroring the behavior observed in Gaussian random matrix ensembles \cite{Jahnke:2025exd}. However, for any finite system size, correlation functions retain a nonzero late-time plateau whose magnitude is set by the system size. To resolve this effect, we analyze the system-size dependence of the full temporal profile of the cumulative OTOC, including its saturation behavior, for both Spin SYK$_2$ and gSpin SYK$_2$ as shown in Fig.\,\ref{fig:finesizecummOTOCggSpinSYK}. We observe that the saturation value of the cumulative OTOC, denoted by $\overline{\mathcal{F}_{p}^{(i,j)}}$, decreases with increasing $N_{\mathrm{Spin}}$, signaling the emergence of asymptotic freeness. Exact freeness is recovered only in the $N_{\mathrm{Spin}} \rightarrow \infty$ limit. This behavior indicates that, at late times, $O_1$ effectively becomes free with respect to its time-evolved counterpart $O_1(t)$, consistent with the predictions of chaotic dynamics, and provides further evidence for the chaotic nature of the (g)Spin SYK models. We have confirmed that this conclusion remains valid for other choices of local operators, such as $O_2 = \sigma_y \otimes \mathbb{I} \otimes \cdots \otimes \mathbb{I}$ and $-i O_1 O_2 = \sigma_z \otimes \mathbb{I} \otimes \cdots \otimes \mathbb{I}$, demonstrating the robustness of the emergent freeness across different operator bases.

In summary, operator growth provides insights that extend beyond the spectral properties of the Hamiltonian, illuminating how the structure of eigenstates influences the dynamics of the system. In the next section, we focus on the eigenstates of the Spin SYK$_2$ model and analyze quantities that are especially relevant to ergodicity and to the resource theory of quantum simulation.

\section{Eigenstates of the Spin SYK Models}
\label{sec:eigprop}

In Section \ref{sec:eigen}, we characterized the chaotic properties of the Hamiltonian via eigenvalue correlations, finding agreement with Gaussian random-matrix statistics. We now turn to the statistical properties of the eigenstates, focusing on a single representative mid-spectrum state, the regime where chaotic behavior is expected to dominate. Although one could instead average over a narrow energy window near the spectral center, the qualitative conclusions remain unchanged. We analyze the ergodic properties of this state in comparison with Haar-random predictions, which serve as benchmarks corresponding to the eigenstates of Gaussian random matrices.

\subsection{Fractal dimensions}
\label{sec:fracdim}

An important way to characterize the properties of an eigenstate---such as its ergodicity or localization properties---is through the fractal dimension \cite{HENTSCHEL1983435}. We follow the convention of Ref.\,\cite{IvanMultifractalRMT}. To define it, consider a many-body eigenstate $|\Psi_{j}\rangle$ expanded in some orthonormal basis ${|\psi_i\rangle}$ as $|\Psi_j \rangle = \sum_{i=1}^d c_i^{(j)}|\psi_i\rangle$ with complex coefficients $c_i^{(j)} = \langle \psi_i |\Psi_j \rangle$ encoding the overlap between the given state and the basis. Here $d$ is the dimension of the Hilbert space, \emph{i.e.}, the number of linearly independent basis vectors. The probability distribution associated with the eigenstate is then given by $|c_i^{(j)}|^2$. One characterizes the distribution by its moments:
\begin{align}
I_{\alpha}(j,d) = \sum_{i=1}^d |c_i^{(j)}|^{2\alpha}\,, \label{momcij}
\end{align}
where $\alpha$ is typically an integer, but can be extended to non-integer values. If the orthonormal basis is chosen to be the computational basis, the coefficients $c_i^{(j)}$ are simply the components of the eigenstate vector. Using \eqref{momcij}, the fractal dimension is defined as \cite{IvanMultifractalRMT}
\begin{align}
D_{\alpha} = \frac{1}{1-\alpha} \frac{1}{\log_2 d} \log_2 I_{\alpha}(j,d) = \frac{1}{N(1-\alpha)} \log_2 I_{\alpha}(j,d)\,, \label{fracdim}
\end{align}
where the second equality holds for $N$-qubit systems with Hilbert space dimension $d = 2^N$. Since we are dealing with spin systems, we use the logarithm with base $2$, although this choice is not essential. It is instructive to examine the behavior of the fractal dimension in two extreme limits. In the fully ergodic phase, all coefficients $c_{i}^{(j)}$ are equal in magnitude and given by the inverse square root of the Hilbert space dimension, namely $c_{i}^{(j)} = 1/\sqrt{d}$. In this case, the moments evaluate to $I_{\alpha}(j,N) = d^{1-\alpha}$, and the fractal dimension is unity, $D_{\alpha} = 1$, for all $\alpha$. Conversely, in the localized phase, where the eigenstate is concentrated on only a few basis states, the moments remain constant with system size. This yields a vanishing fractal dimension, $D_{\alpha} = 0$ in the large $N$ limit. Between these limits, as realized for example in certain Rosenzweig–Porter–type matrix models, eigenstates can be multifractal \cite{Luitz_Multifractal, Khaymovich:2021tkj}, \emph{i.e.}, non-trivial dependence on $\alpha$, and the fractal dimension takes values strictly between zero and one: $0 < D_{\alpha} < 1$.

\begin{figure}
    \centering
\includegraphics[width=0.6\linewidth]{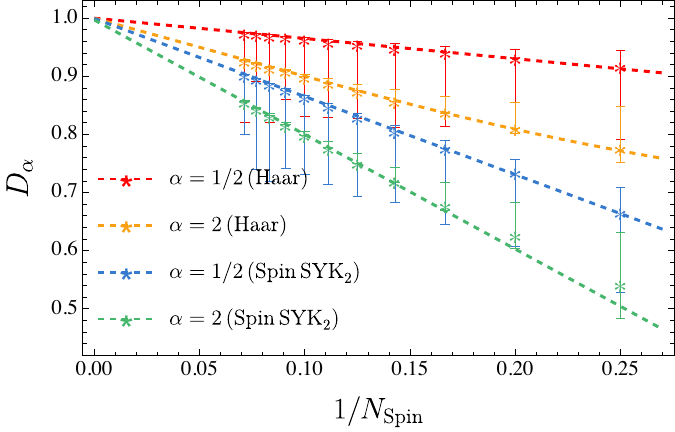}
    \caption{The fractal dimension $D_{\alpha}$ for a single mid-spectrum eigenstate of the Spin SYK$_2$ model as a function of $1/N_{\mathrm{Spin}}$ (the system size corresponds to $N_{\mathrm{Spin}} =4$ to $N_{\mathrm{Spin}} = 14$) for two different values, $\alpha =1/2$ and $\alpha =2$, shown by the blue and the green stars, with the numerical fitting given by the corresponding dashed lines. The red and orange stars represent the numerical results for Haar-random (GUE) values, with the dashed lines indicating the corresponding analytic results \eqref{DGUE} for $\alpha = 1/2$ and $\alpha = 2$, respectively. For each point, the error bars are shown in respective colors. In all cases, averages are taken over $2^{16-N_{\mathrm{Spin}}}$ Hamiltonian samples.}
  \label{fig:DalphabyNallplotMidES}
\end{figure}

For Haar-random states, constructed from the eigenvectors of the GUE, one can obtain an analytic expression for the fractal dimension. In this case, the coefficients of the eigenvectors follow the Porter–Thomas distribution \cite{PorterThomas}, from which the moments can be computed directly. One finds \cite{IvanMultifractalRMT}
\begin{align}
    I_{\alpha}^{\mathrm{GUE}} (d) = \frac{\alpha! \, d!}{(d-1+\alpha)!}\,,
\end{align}
and by substituting this into \eqref{fracdim}, the fractal dimension is readily obtained as \cite{IvanMultifractalRMT}
\begin{align}
    D_{\alpha}^{\mathrm{GUE}} = \frac{1}{1-\alpha} \frac{1}{\log_2 d} \log_2 \left[\frac{\alpha! \, d!}{(d-1+\alpha)!}\right]\,, \label{DGUE}
\end{align}
with $d = 2^N$. The GOE case can be derived similarly \cite{IvanMultifractalRMT}, although it leads to expressions different from \eqref{DGUE}. Likewise, the limit $\alpha \to 1$ can be obtained using the standard procedure, as is commonly employed in entanglement entropy calculations.

Figure \ref{fig:DalphabyNallplotMidES} illustrates the fractal dimension $D_{\alpha}$ for $\alpha = \{1/2, 2\}$, computed for a single representative mid-spectrum eigenstate at various system sizes $N \equiv N_{\mathrm{Spin}}$. Blue and green stars denote the results for $\alpha = 1/2$ and $\alpha = 2$, respectively, with the corresponding dashed lines representing numerical fits. For comparison, Haar-random results are shown by red and orange stars, with dashed lines given by the analytical predictions \eqref{DGUE}. At finite $N_{\mathrm{Spin}}$, the Spin SYK model exhibits a clear deviation from Haar-random values, reflecting its fractal eigenstate structure. This deviation diminishes as $N_{\mathrm{Spin}}$ increases, with $D_{\alpha} \to 1$ in the $N_{\mathrm{Spin}} \to \infty$ limit, where the fractal dimension coincides with the Haar prediction and the eigenstates become fully ergodic. Such finite-size deviations are commonly attributed to the ``weakly ergodic'' character of non-integrable many-body systems \cite{IvanMultifractalRMT}, and in the Spin SYK case, the local structure of the constituent bosons provides an additional constraint that naturally explains these deviations.

\subsection{Stabilizer R\'enyi entropy}
\label{sec:ren}

Next, we turn to the characterization of the stabilizer properties of the eigenstates. The notion of \emph{magic}, or non-stabilizerness, has recently emerged as a key measure of the quantum resources required for universal quantum computation \cite{Gottesman:1997zz, Aaronson:2004xuh}. Stabilizer states---eigenstates of Pauli operators generating a stabilizer group---are classically simulable and thus carry no magic, whereas generic many-body eigenstates typically exhibit significant deviations from this stabilizer structure. A convenient measure of this deviation is the Stabilizer R\'enyi Entropy (SRE) \cite{Leone:2021rzd}. For an $N$-qubit pure state $|\psi\rangle$, the order-$\alpha$ SRE is defined as
\begin{align}
\mathcal{M}_{\alpha}(|\psi\rangle) = \frac{1}{1-\alpha}\, \log_2 \sum_{P \in \mathcal{P}_N} \frac{1}{2^N}\, \big|\langle \psi | P | \psi \rangle \big|^{2\alpha} \,. \label{SRE}
\end{align}
Here, $\mathcal{P}_N$ denotes the $N$-qubit Pauli group, generated by tensor products of single-qubit Pauli operators $\{\mathbb{I}, \sigma^x, \sigma^y, \sigma^z\}$. The parameter $\alpha$ sets the order of the SRE, with the regime $\alpha \ge 2$ exhibiting good monotonicity properties \cite{Leone:2021rzd}. For a generic state $|\psi\rangle$, the SRE quantifies its deviation from stabilizer states, which by definition yield zero SRE. In contrast, Haar-random states display nearly uniform support over the Pauli basis. For arbitrary system size and $\alpha$, the SRE of Haar-random states  (GUE) admits an analytic closed-form expression \cite{Turkeshi:2023lqu}:
\begin{align}
\mathcal{M}_{\alpha}\big|^{\mathrm{Haar}}_{\mathrm{GUE}} &= \frac{1}{1-\alpha}\log_2 \left(\frac{1}{d} + \frac{2(d-1) \Gamma(\alpha + \frac{1}{2}) \Gamma(\frac{d+3}{2})}{d \sqrt{\pi}\, \Gamma(\alpha+\frac{d+1}{2})}\right)\,, \label{GUESREq}
\end{align}
where $d=2^N$ is the dimension of $N$ qubits. The analogous case of GOE can also be obtained in analytic form \cite{Turkeshi:2023lqu}. The $\alpha = 2$ case is the most-studied example, for which the above expression simplifies to \cite{Turkeshi:2023lqu, Odavic:2024lqq}
\begin{align}
\mathcal{M}_2\big|^{\mathrm{Haar}}_{\mathrm{GUE}} = - \log_2 \left(\frac{4}{3 + 2^{N}}\right)\,.\label{specM}
\end{align}
To evaluate the SRE, we numerically diagonalize the Hamiltonian, extract eigenstates, and consider a single representative mid-spectrum eigenstate. We compute the SRE \eqref{SRE} for $\alpha = 2$ and $\alpha = 3$ for the mid-spectrum eigenstate of the Spin-SYK$_2$ model. Since the model belongs to the GUE universality class, in the large-$N$ limit, \eqref{GUESREq} yields
\begin{align}
    \lim_{N \rightarrow \infty} \frac{1}{N} \,\mathcal{M}_{\alpha}\big|^{\mathrm{Haar}}_{\mathrm{GUE}} = \frac{1}{\alpha - 1}\,. \label{SREinfN}
\end{align}
For $\alpha =2$, this limiting value approaches unity, also apparent from \eqref{specM}.
\begin{figure}
    \centering
\includegraphics[width=0.6\linewidth]{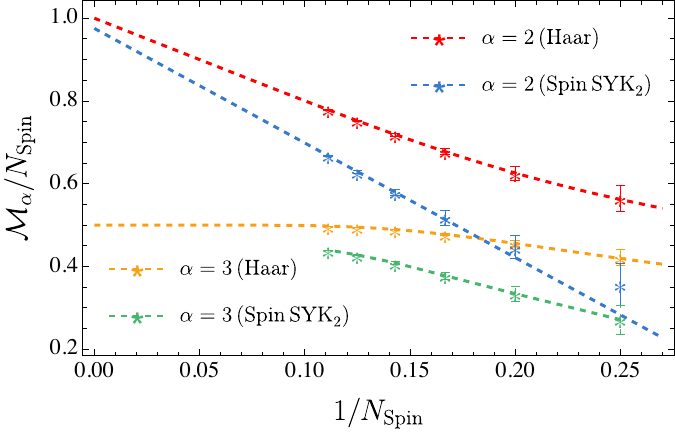}
    \caption{The Stabilizer R\'enyi entropy  $\mathcal{M}_{\alpha}$, rescaled by the number of spins, for a single mid-spectrum eigenstate of the Spin SYK$_2$ model as a function of $1/N_{\mathrm{Spin}}$ for two different values, $\alpha =2$ and $\alpha =3$, shown by the blue and the green stars, with the numerical fitting given by the corresponding dashed lines. The red and orange stars represent the numerical results for Haar-random (GUE) values, with the dashed lines indicating the corresponding analytic results for $\alpha = 2$ and $\alpha = 3$, respectively. The error bars are shown. In all cases, averages are taken over $2^{16-N_{\mathrm{Spin}}}$ Hamiltonian samples.}
    \label{fig:M2byNallplotMidES}
\end{figure}

Figure \ref{fig:M2byNallplotMidES} shows the dependence of the $\alpha = {2,3}$ SRE as a function of the inverse system size $1/N_{\mathrm{Spin}}$. The Spin SYK$_2$ results, averaged over $2^{16-N{\mathrm{Spin}}}$ Hamiltonian samples, are shown as blue and green stars for $\alpha = 2$ and $\alpha = 3$, respectively. The corresponding dashed lines indicate the numerical fits. For comparison, the red and orange stars represent the numerical results for the Haar-random (GUE) predictions. The red and orange dashed lines are given by \eqref{GUESREq} (the red one is \eqref{specM} in particular), with the limiting value as $1/N_{\mathrm{Spin}} \rightarrow 0$ given by \eqref{SREinfN}.

Deviations from the Haar-random predictions are visible for generic $\alpha$, but they decrease as $N_{\mathrm{Spin}}$ becomes large. This behavior is fully consistent with the fractal dimensions $D_{\alpha}$: mid-spectrum eigenstates are ``weakly ergodic'' \cite{IvanMultifractalRMT}, reflecting the local interactions present in the Spin SYK$_2$ model. Such local interactions also slow down the entropy production compared to the other SYK variants \cite{Tezuka:2022mrr}, as recently shown in Ref.\,\cite{Pathak:2025udi} for the Spin SYK models with quartic interactions. Analysis of the gSpin-SYK$_2$ model indicates a qualitatively analogous behavior.

\section{Conclusion and Outlook}\label{sec:con}

In this work, we uncover quantum chaotic behavior in one of the simplest quadratic SYK-like models. We show that this model exhibits hallmark signatures of quantum chaos, including gap-ratio statistics consistent with RMT, a ramp–plateau structure in the spectral form factor, and a characteristic peak in Krylov spread complexity. Operator growth in Krylov space aligns with the operator-growth hypothesis \cite{Parker:2018yvk}, while the late-time decay of cumulative OTOCs \cite{Jahnke:2025exd} signals the emergence of \emph{asymptotic freeness} from the perspective of free probability. This class of quadratic SYK-like models thus emerges as one of the simplest and most accessible candidates for realizing quantum chaotic dynamics on future quantum computers \cite{Hanada:2023rkf, Asaduzzaman:2023wtd}.

An intriguing direction for future work is to extend the analysis to include all three Pauli matrices, as well as higher-dimensional qudit generalizations based on $SU(d)$, $SO(d)$, and $Sp(d)$ matrix models \cite{Hanada:2025pis}. A key question is whether these more complex systems preserve the quadratic chaos behavior observed in simpler variants. Initial investigations \cite{Hanada:2025pis} of spectral statistics suggest that the answer is affirmative. Our Hamiltonian is fully \emph{dense}, and it would be interesting to sparsify the model to determine whether the chaotic behavior persists. In particular, it remains an open question whether the critical sparsity threshold scales with system size in a similar power-law fashion \cite{Garcia-Garcia:2020cdo, Orman:2024mpw, Nandy:2024wwv}. Generalizing the model to non-Hermitian cases, similar to the fermionic SYK model \cite{Garcia-Garcia:2021rle, Nandy:2024wwv}, and to open systems using Lindbladian dynamics \cite{Kulkarni:2021gtt, Sa:2021tdr, Bhattacharjee:2022lzy, Bhattacharjee:2023uwx}, represents another promising future direction.

More broadly, it is important to clarify how quadratic (or more generally, small-$q$) models differ from large-$q$ models with emergent gravitational duals. A key question is whether these simpler models display maximally chaotic behavior at finite temperature, saturating the chaos bound \cite{Maldacena:2015waa}. Our model may share certain features with spin glasses \cite{Bray_1980, Anous:2021eqj} or other quadratic models studied in contexts such as quantum batteries (Sec.\,5 and Appendix C of Ref.\,\cite{Rosa:2019jin}). The non-integrable behavior of these seemingly simple quadratic models arises because their hard-core boson constituents introduce interactions, leading to genuine many-body chaos. In this context, developing refined measures of complexity capable of distinguishing quadratic two-body chaos from true many-body chaos \cite{Andreanov:2023nxk, Flynn:2024thf} would be especially valuable.

From the perspective of eigenstates, through the scaling of fractal dimensions and SRE, our results demonstrate that mid-spectrum eigenstates of the quadratic models fail to reach the Haar-random prediction at any finite $N_{\mathrm{Spin}}$, approaching it only in the infinite-size limit. This behavior characterizes their weakly ergodic nature. Together with the observation of a vanishing cumulative OTOC, this suggests that fully ergodic eigenvectors are not necessary for the emergence of freeness. It would be interesting to contrast this with systems whose eigenvectors are fully ergodic: does approximate freeness appear at larger sizes (or later timescale) when eigenvectors are only weakly ergodic compared with the fully ergodic regime \cite{Jahnke:2025exd}? A natural direction for addressing this question is to examine higher-dimensional qudit SYK models, which may host more ergodic-like eigenstates and permit analytic insight from Eq.\,\eqref{GUESREq}. Moreover, due to the local string structure of constituents, the behavior of ground states can differ significantly from that of mid-spectrum states. Recent studies have taken promising steps toward classifying the stabilizerness of both ground states and mid-spectrum states in the fermionic SYK model across the chaotic–integrable transition \cite{Jasser:2025myz, Santra:2025pvn}.

Interestingly, holographic states also possess substantial magic \cite{White:2020zoz}, making it worthwhile to investigate whether quenches of thermofield-like states in Spin SYK models, analogous to the analysis in Ref.\,\cite{Goto:2021anl}, exhibit any nontrivial behavior in magic. Whether such properties can be leveraged to demonstrate the size-winding mechanism and thereby enable reliable teleportation protocols \cite{Brown:2019hmk} remains an outstanding open question, and a systematic exploration of these features within the framework of resource theory is left for future work.

\section{Acknowledgements}

We acknowledge A. Bhattacharya for initial collaboration, and X. Cao, V. Jahnke, A. Joseph, I. M. Khaymovich, B. Madlala, D. Rosa, M. Tezuka, and N. Zenoni for insightful discussions and valuable feedback on the draft. We particularly thank M. Tezuka for sharing some unpublished results on Spin SYK models. The work of P.N. is supported by the JSPS Grant-in-Aid for Transformative Research Areas (A) ``Extreme Universe'' No. 21H05190, FWO-Vlaanderen projects G012222N and G0A2226N, and by the VUB Research Council through the Strategic Research Program in High-Energy Physics.

\appendix

\section{Krylov Complexities}\label{app:krycomplex}

We present a concise introduction to Krylov complexity, with more detailed discussions available in Refs.\,\cite{Nandy:2024evd, Rabinovici:2025otw}. Krylov complexity \cite{Parker:2018yvk, Balasubramanian:2022tpr} provides a systematic framework to characterize the spreading of quantum states and operators under time evolution. For states, one considers the unitary evolution of an initial state, while for operators, one follows the Heisenberg evolution of an initially localized  operator:
\begin{align}
    \ket{\Psi(t)} &= e^{-iHt}\ket{\Psi_0}\,,\,~~~~~~ (\mathrm{state~evolution})\,,\\ O(t) &= e^{iHt}\,O_0\,e^{-iHt}\,, ~~~(\mathrm{operator~evolution})\,.
\end{align}
Here $H$ is the time-independent Hamiltonian, $\ket{\Psi_0}$ and $O_0$ denote the initial state and Hermitian operator, respectively. We exclude cases where $\ket{\Psi_0}$ is an eigenstate of $H$, or $O_0$ is a conserved operator, since in such cases no nontrivial evolution occurs.

The essential idea of Krylov complexity is to express the evolving state or operator in the Krylov basis, constructed iteratively through successive applications of the Hamiltonian (for states) or Liouvillian (for operators) $\mathcal{L}(\cdot) = [H,\cdot]$, followed by a Gram-Schmidt like orthogonalization (re-orthogonalization \cite{Rabinovici:2020ryf} in some cases for numerical stability). On this basis, the time-evolved state/operator takes the following form:
\begin{align}
    \ket{\Psi(t)} &= \sum_{n=0}^\infty \psi_n(t)\,\ket{K_n}\,,~~~ (\mathrm{state~evolution})\,, \\
   |O(t)) &= \sum_{n=0}^\infty i^n \varphi_n(t)\,|\mathcal{O}_n)\,, ~~~~(\mathrm{operator~evolution})\,,
\end{align}
with an appropriate choice of inner product in each case \cite{viswanath1994recursion}. Here $\ket{K_n}$ and $|\mathcal{O}_n)$ are the orthonormal Krylov basis for the states and the operators, with the initial condition $\ket{K_n} = \ket{\psi_0}$ and $|\mathcal{O}_0) = |O_0)$. The functions $\psi_n(t)$ and $\varphi_n(t)$ are known as Krylov basis wavefunctions and are related to the corresponding Lanczos coefficients (different for states or operators) through the three-term recursion relations:
\begin{align}
i \partial_t \psi_n(t) &:= b_n \psi_{n-1} + a_n \psi_n + b_{n+1} \psi_{n+1} \,,~~~(\mathrm{state~evolution})\,,\\
\partial_t \varphi_n(t) &:= b_n \varphi_{n-1} -  b_{n+1} \varphi_{n+1} \,, ~~~~~~~~~~~~~ (\mathrm{operator~evolution})\,.
\end{align}
Hence, in this representation, the Schr\"odinger or Heisenberg equations of motion reduce to a one-dimensional particle-hopping problem along the Krylov chain. Krylov complexity is then defined as the mean position of the particle in this chain:
\begin{align}
K_{S}(t) &:= \sum_{n} n|\phi_n(t)|^2 \,,~~~(\mathrm{state~complexity})\,,\\
K_{O}(t) &:= \sum_{n} n|\varphi_n(t)|^2 \,, ~~~ (\mathrm{operator~complexity})\,.
\end{align}
The summation is taken up to the Krylov dimension, which is bounded by the Hilbert space dimension (for states) \cite{Balasubramanian:2022tpr} or the operator space dimension (for operators) \cite{Rabinovici:2020ryf}. For operators, we omit the subscript and denote the Krylov complexity simply as $K(t)$. The structure of the Krylov chain is determined by the Lanczos coefficients, which in turn control the growth of complexity. In chaotic systems, the complexities exhibit characteristic behavior: operator complexity typically shows an early stage of exponential growth (consequence of the linear growth of Lanczos coefficients $b_n$), while state complexity develops a pronounced peak before saturation. It is important to note that the operator growth hypothesis—that $b_n$ grows linearly—applies only to operator complexity. By contrast, integrable systems show non-universal behavior, with the growth of complexity depending sensitively on model-specific details, except in a few pathological cases \cite{Dymarsky:2021bjq, Bhattacharjee:2022vlt}.

\begin{figure}
    \centering
\includegraphics[width=0.57\linewidth]{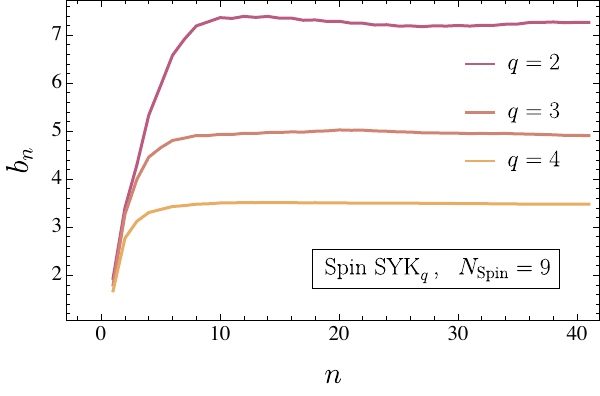}
    \caption{The behavior of Lanczos coefficients $b_n$ for spin SYK$_q$ model with $q=2,3,4$. The averages were taken on 50 samples of the Hamiltonian.}
\label{fig:plotbnSpinSYKallq}
\end{figure}

We briefly highlight the algorithm employed for the operators. More details, including the algorithms for the operator and the states, can be found in \cite{Nandy:2024evd, Rabinovici:2025otw}.
We start with initiating the normalized operator $|\mathcal{O}_0) = |O_0)/(O_0|O_0)$ and $|\mathcal{O}_{-1}) = b_0 := 0$, followed by the steps for $n \geq 1$:
\begin{align}
    &1.~ a_{n-1} = (O_{n-1}| \mathcal{L}|O_{n-1}) \nonumber\\
    &2.~ |A_n) = \mathcal{L}|\mathcal{O}_{n-1}) -a_{n-1} |\mathcal{O}_{n-1}) - b_{n-1} |\mathcal{O}_{n-2}) \,. \nonumber 
\\
&3.~ \mathrm{If}~ (A_n|A_n) = 0, ~\mathrm{Stop}. ~\mathrm{Otherwise~compute}~ b_n = \sqrt{(A_n|A_n)}\,. \nonumber\\
&4.~ |\mathcal{O}_n) = \frac{1}{b_n}|A_n)\,~~\mathrm{and}~~ a_n = (\mathcal{O}_n|\mathcal{L}|\mathcal{O}_n)\,. \nonumber
\end{align}
This algorithm has been employed to generate the Fig.\,\ref{fig:plotbnggSpinSYK2}, Fig.\,\ref{fig:plotbnSpinSYK2diffO}, and Fig.\,\ref{fig:plotbnSpinSYKallq}. However, due to the iterative nature of the Lanczos algorithm, it is susceptible to numerical instabilities, as small errors can rapidly propagate through the iterations. Consequently, while the lower-order (small $n$) Lanczos coefficients can be determined accurately, the higher-order (large $n$) coefficients may become unreliable. These instabilities can be mitigated using several strategies. One approach is to increase the numerical precision, while another is to explicitly re-orthogonalize the Krylov vectors at each iteration step \cite{Rabinovici:2020ryf}. This strategy underlies the Arnoldi method \cite{Bhattacharya:2022gbz, Bhattacharjee:2022lzy, Nandy:2024evd}, which generalizes the above algorithm for when $\mathcal{L}$ is non-Hermitian, and explicitly takes care of the re-orthogonalization method. We employ the Arnoldi iteration in this work, especially to compute the entire Lanczos sequence in Fig.\,\ref{fig:bnKryplotggSpinSYK2N6}. The algorithm is as follows:
We initialize with a normalized vector $|\mathcal{V}_0) =  |O_0)/(O_0|O_0)$. Then, for $k=1, 2, \ldots,n$, and $j = 0,\cdots,k-1$, an iterative construction yields the following:
\begin{align}
   &1.~|\mathcal{U}_k )= \mathcal{L}^{\dagger} \,|\mathcal{V}_{k-1})\,. \nonumber \\
    &2. ~h_{j,k-1} = ( \mathcal{V}_j|\mathcal{U}_k )\,.~~~~~~~~~~~~~~~~~~~~~~~~ ~~~~~~~~~~~~ ~~~~~~~~~~~~~~~~~~~~~~~~~   \nonumber \\
    &3. ~|\tilde{\mathcal{U}}_{k} ) =|\mathcal{U}_{k} )-\sum_{j=0}^{k-1} h_{j,k-1} |\mathcal{V}_{j} )\,. \nonumber \\
    &4. ~ h_{k,k-1} =  \sqrt{({\tilde{\mathcal{U}}_k|\tilde{\mathcal{U}}_k})}\,. \nonumber \\
    &5. ~ \mathrm{If}~h_{k,k-1}=0,~ \mathrm{Stop}.~\mathrm{Otherwise,~define~}|\mathcal{V}_k )= \frac{|\tilde{\mathcal{U}}_k )}{h_{k,k-1}}\,. \nonumber
\end{align}
For unitary evolution, the Liouvillian is Hermitian $\mathcal{L}^{\dagger} = \mathcal{L}$, and it takes a tridiagonal structure with coefficients $h_{m,n} = (\mathcal{V}_m|\mathcal{L}^{\dagger}|\mathcal{V}_n)$. The Arnoldi basis $|\mathcal{V}_n)$ becomes (up to constant factors or linear combinations) the associated Krylov vectors $|\mathcal{O}_n)$, and the Lanczos coefficients are given by $a_n = h_{n,n}$ and $b_n = h_{n,n-1}$. Alternative approaches, such as partial re-orthogonalization strategies \cite{Rabinovici:2020ryf, Rabinovici:2025otw} of the Krylov vectors, can also be used to control numerical errors. We choose the initial operators as in \eqref{inop1} and apply the algorithm to the full Hamiltonian, without resolving symmetry sectors. While such a resolution is essential for analyzing spectral properties, it is not required for the operator dynamics studied here.

\section{Spin SYK models with $q \geq 2$} \label{AppB}

\begin{figure}
    \hspace*{-0.3 cm}
\includegraphics[width=1\linewidth]{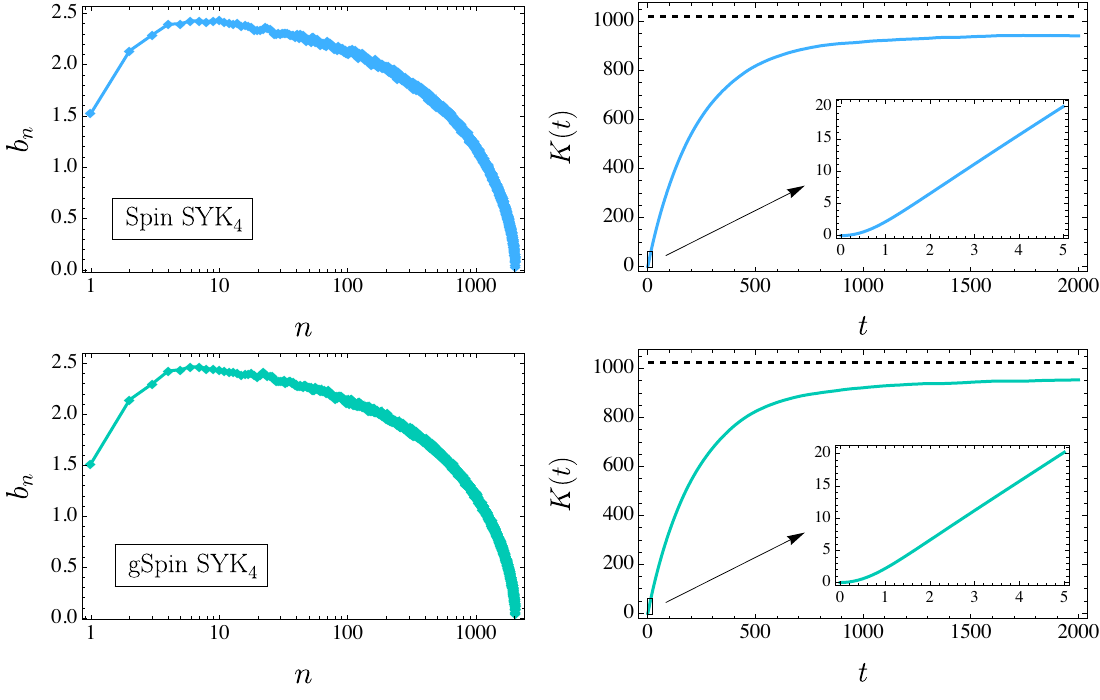}
    \caption{Growth of Lanczos coefficients $b_n$ and Krylov complexity for Spin SYK$_4$ (upper panel) and gSpin SYK$_4$ (lower panel). The Krylov space dimension is $K_{\mathrm{dim}} = 2040 + O(1)$ for either case, corresponding to the maximal saturation of complexity $K_{\mathrm{dim}}/2$, shown by the black dashed line in the right figures. The inset shows the early-time behavior. We take $N = 6$ with $20$ realizations of Hamiltonians.}
\label{fig:bnKryplotggSpinSYK4L6}
\end{figure}

In this Appendix, we discuss the growth of the Lanczos coefficients for $q \geq 2$ in the Spin SYK models. Figure \ref{fig:plotbnSpinSYKallq} illustrates the behavior of the Lanczos coefficients $b_n$ for different values $q = 2, 3, 4$ at a fixed system size $N = 9$, averaged over 50 ensembles of the Hamiltonian \eqref{SpinSYKq}. The observed saturation arises from the finite size of the system. We also note that as $q$ increases, the saturation level decreases, although this decrease is generally not linear. The entire spectrum of Lanczos coefficients and Krylov complexity for (g)Spin SYK$_4$ is shown in Fig.\,\ref{fig:bnKryplotggSpinSYK4L6}.

\section{Some simpler versions of the Hamiltonian} \label{app:simpham}
Our starting point of the analysis was the Hamiltonian \eqref{SpinSYKq}, which consists of two types of terms:  
\begin{enumerate}
    \item Interactions involving only $\sigma_x$ (or only $\sigma_y$) on the same or different sites (genuine two-body),  
    \item Cross interactions between $\sigma_x$ and $\sigma_y$ on the same or different sites (genuine two-body).  
\end{enumerate}
A natural question arises: what are the minimal ingredients at the quadratic level necessary to generate chaos? In other words, is any single type of term sufficient, or are all of them required? To answer this, we examine each case separately.

\begin{figure}
    \centering
    \includegraphics[width=1\linewidth]{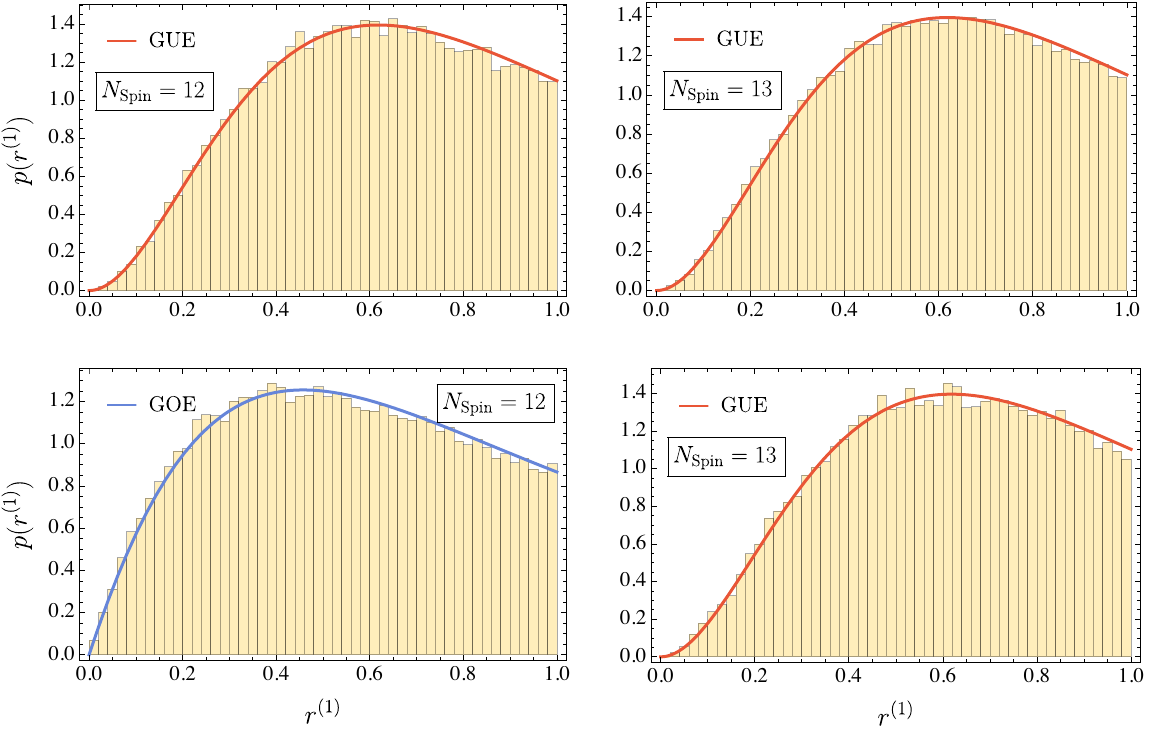}
    \caption{The distribution of level-spacing ratios of the eigenvalues of the Hamiltonian \eqref{SpinSYKq2xy}, including the same site (top panel, different $N_{\mathrm{Spin}}$), and excluding (genuine two-body) the same site (bottom panel, different $N_{\mathrm{Spin}}$). When the same-site interactions are included, the universality class is GUE for any $N_{\mathrm{Spin}}$, whereas for the \emph{genuine} two-body case, even $N_{\mathrm{Spin}}$ odd $N_{\mathrm{Spin}}$ correspond to the GOE and GUE classes. In all cases, a total $2^{18- N_{\mathrm{Spin}}}$ Hamiltonian samples are taken.}
    \label{fig:HistspecvalggHxyplot}
\end{figure}

We now demonstrate that the Spin–SYK Hamiltonians admit further simplification. We restrict attention to a simple subsector of the Hamiltonian defined in \eqref{SpinSYKq} by considering the $\sigma_x$-$\sigma_y$ interaction:
\begin{align}
H_{\mathrm{Spin\,SYK}_2}^{xy} = \sqrt{\frac{1}{2N_{\mathrm{Spin}}}}  \sum_{1 \leq i_1 < i_2 \leq 2N_{\mathrm{Spin}}} i^{\eta_{i_1 i_2}}\, \xi_{i_1 i_2}\, J_{i_1 i_2} \,  O_{i_1} O_{i_2}\,. \label{SpinSYKq2xy}
\end{align}
Here, the additional $\xi_{i_1 i_2}$ term removes the $\sigma_x-\sigma_x$ or $\sigma_y-\sigma_y$ terms from the Hamiltonian. Due to the presence of interactions between $\sigma_x$ and $\sigma_y$, the only commuting operator is $\Gamma_z$, which allows the Hamiltonian to be written in a block-diagonal form with two blocks corresponding to eigenvalues $\pm 1$. In Fig.\,\ref{fig:HistspecvalggHxyplot}, we show the level-spacing ratio distribution for a single block of this Hamiltonian, which clearly exhibits signatures of quantum chaos. Moreover, the universality class depends on the parity of $N_{\mathrm{Spin}}$: when same-site interactions are included, the universality class belongs to GUE for all $N_{\mathrm{Spin}}$, whereas for the \emph{genuine} two-body case, it is GOE for even $N_{\mathrm{Spin}}$ and GUE for odd $N_{\mathrm{Spin}}$.

\bibliography{bibliography.bib}
\bibliographystyle{JHEP.bst}

\end{document}